\documentclass[journal=ancac3,manuscript=article]{achemso}

\usepackage[version=3]{mhchem} 
\usepackage{textcomp, gensymb}
\usepackage{xcolor}
\usepackage{float}

\usepackage{float} 
\usepackage{array} 
\usepackage{caption}
\usepackage{subcaption}
\usepackage{graphicx}
\usepackage{seqsplit}
\author{Elisa Ballin}
\affiliation{Dipartimento di Fisica, Sapienza Università di Roma, Piazzale A. Moro 5, 00185 Roma, Italy}
\alsoaffiliation{Istituto dei Sistemi Complessi, Consiglio Nazionale delle Ricerche, Piazzale A. Moro 5, 00185 Roma, Italy}
\email{elisa.ballin@uniroma1.it}
\author{Francesco Brasili}
\affiliation{Istituto dei Sistemi Complessi, Consiglio Nazionale delle Ricerche, Piazzale A. Moro 5, 00185 Roma, Italy}
\alsoaffiliation{Dipartimento di Fisica, Sapienza Università di Roma, Piazzale A. Moro 5, 00185 Roma, Italy}
\author{Michael Sztucki}
\affiliation{ESRF, The European Synchrotron, 71 avenue des Martyrs CS40220, 38043 Grenoble Cedex 9, France}
\author{Lorenzo Rovigatti}
\affiliation{Dipartimento di Fisica, Sapienza Università di Roma, Piazzale A. Moro 5, 00185 Roma, Italy}
\author{Simona Sennato}
\affiliation{Istituto dei Sistemi Complessi, Consiglio Nazionale delle Ricerche, Piazzale A. Moro 5, 00185 Roma, Italy}
\alsoaffiliation{Dipartimento di Fisica, Sapienza Università di Roma, Piazzale A. Moro 5, 00185 Roma, Italy}
\author{Emanuela Zaccarelli}
\affiliation{Istituto dei Sistemi Complessi, Consiglio Nazionale delle Ricerche, Piazzale A. Moro 5, 00185 Roma, Italy}
\alsoaffiliation{Dipartimento di Fisica, Sapienza Università di Roma, Piazzale A. Moro 5, 00185 Roma, Italy}
\email{emanuela.zaccarelli@cnr.it}

\title[]{Using fast-reactive crosslinkers to modulate the internal structure of thermoresponsive microgels}
\abbreviations{DLS, SAXS, PNIPAM, BIS, EGDMA}

\begin{document}

\begin{figure}[H]
\centering
\includegraphics[width=8.25cm,height=4.45cm,keepaspectratio]{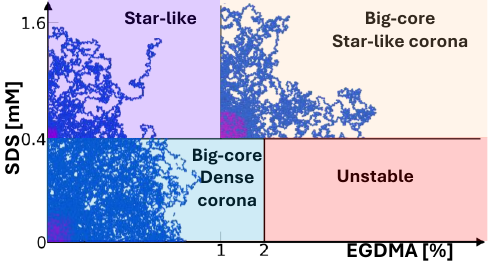}
\label{For Table of Contents Only}
\end{figure}
\begin{abstract}

\noindent The internal architecture of poly(N-isopropylacrylamide) (PNIPAM) microgels, which switches from  fuzzy-sphere to star-like when the standard \seqsplit{N,N'-methylenebis(acrylamide)} (BIS) crosslinker is replaced with ethylene glycol dimethacrylate (EGDMA), critically determines their interactions and swelling behavior.
Here, we systematically investigate the role of the surfactant and crosslinker content in modulating the internal structure of the microgels using Dynamic Light Scattering, Small-angle X-ray Scattering and monomer-resolved numerical simulations. We reveal that the presence of the surfactant is crucial for obtaining the star-like architecture, and that the transition from the star-like regime to a more core-dominated structure occurs above a threshold EGDMA concentration.
Monomer-resolved simulations capture how the role of surfactant differs between EGDMA-crosslinked and BIS-crosslinked microgels. Our findings establish a direct synthesis–structure relationship, providing a clear guidance for the rational design of soft, star-like microgels with ultra-soft interactions, strenghtening the connection between microgels and model star polymers.
\end{abstract}

\section{Introduction}
 
\noindent Colloidal microgels, internally made by a crosslinked polymer network, have a typical size ranging from 50 nm to 5 $\mu$m~\cite{pelton2000temperature}. This can be tuned in many ways, for example by adjusting the affinity between the solvent in which they are suspended, by varying the polymer composition, and, also, by adding stabilizing surfactants to the polymerization process taking place during the synthesis~\cite{fernandez2011microgel}. 
The ability of microgels to respond to external stimuli, most notably temperature for poly(N-isopropylacrylamide) (PNIPAM) ones, has made them valuable model systems in soft matter fundamental research~\cite{lyonPolymerColloidDuality2012a,yunker2014physics,brijitta2019responsive,scotti2022softness} and promising platforms for applications ranging from drug delivery to sensing and catalysis~\cite{plamperFunctionalMicrogelsMicrogel2017b,kargNanogelsMicrogelsModel2019b}.
The internal structure of microgels is largely determined by the choice of synthesis conditions. In particular,  PNIPAM microgels are conventionally crosslinked with N,N'-methylenebisacrylamide (BIS), yielding an internal structure that is well-described by the fuzzy-sphere model, featuring a densely crosslinked core and a diffuse, loosely crosslinked corona. This morphology arises from the differing reactivity ratios of NIPAM monomers and BIS crosslinkers during precipitation polymerization, which leads to a preferential accumulation of BIS within the core. 
Several investigations have sought to go beyond this conventional architecture, attempting to engineer microgels with tailored internal structures. These include more homogeneous microgels starting from oligomers~\cite{mueller2018dynamically}, ultralow-crosslinked (ULC) microgels that rely on self-crosslinking of NIPAM~\cite{gao2003cross,bachman2015ultrasoft,hazraStructureResponsiveMicrogels2024a}, as well as hollow microgels, consisting of a polymer shell with an empty cavity~\cite{nayakHollowThermoresponsiveMicrogels2005a}. Additionally, core-shell microgels comprising a rigid core onto which a PNIPAM shell is grafted are also very popular.~\cite{luThermosensitiveCoreShell2011a}
In this context, recent works have put forward the hypothesis that PNIPAM microgels crosslinked with ethylene glycol dimethacrylate (EGDMA) can exhibit a star polymer architecture~\cite{rivas2022link,ruiz-francoConcentrationTemperatureDependent2023c,vialetto2025controlling}, characterized by multiple linear arms radiating from a central core. This hypothesis was confirmed in Ref.~\cite{ballinStarLikeThermoresponsiveMicrogels2025}, which reported a direct comparison between measured form factors for these microgels and numerical simulations of star polymers and star-like microgels. This work thus bridged the gap between star polymers and microgels, combining the ultrasoft nature of the interactions of the former~\cite{likos1998star} with the facile synthesis and thermoresponsiveness of the latter.
The mechanism underlying this important result is rather simple: EGDMA polymerizes much faster than NIPAM and also considerably faster than BIS, so that it tends to exclusively, rather than preferentially, accumulate within a tiny central core inside the particle. 
However, Ref.~\citenum{ballinStarLikeThermoresponsiveMicrogels2025} only focused on EGDMA-crosslinked microgels synthesized by precipitation polymerization in the presence of sodium dodecyl sulfate (SDS) surfactant at two molar fractions ($c$, mol\%) of EGDMA.
In these conditions, the internal structure of the particles at low crosslinker density ($c=1$\%) was shown to be fully compatible with that of a star polymer, thanks to a direct comparison with a numerical model of star polymers. In addition, the effective interactions between these star-like microgels, calculated by numerical simulations, were found to be ultrasoft~\cite{papettiStarlikeMicrogelsVs2026}, in agreement with expectations for star polymers~\cite{likos1998star,likos2006soft}.
However, at high crosslinker content ($c=10$\%), the size of the EGDMA-rich core is no longer negligible with respect to the total size of the particle, so that the form factor is no longer compatible with that of star-like particles. To model these features, the so-called core-fuzzy-shell model was employed, successfully reproducing the form factors for $c=10$\%, which reduces to the star-polymer model introduced by Dozier et al.~\cite{dozierColloidalNatureStar1991} in the limit of a point-like core for $c=1$\%. Moreover, EGDMA-crosslinked microgels were found to display a superior swelling ability and a sharper volume phase transition (VPT) as compared to standard BIS-crosslinked microgels, confirming the pioneering findings by Kratz and coworkers~\cite{kratzVolumeTransitionStructure2002}. After these initial reports, there is still the need to perform a precise and systematic assessment of the relationship between synthesis protocol and structure of the polymer network to clarify under which preparation conditions the star-like regime can be recovered. To this aim, in this work we independently explore the role of surfactant added in the synthesis, which is usually employed to vary the size of standard microgels and of the crosslinker molar fraction. Both investigations are needed in order to establish the optimal conditions for obtaining star-like microgels and to fully exploit the potential offered by soft particles capable of combining the properties of star polymers and microgels.\\
\noindent Regarding the role of surfactant, it is important to remark that it is well-established that, for BIS-crosslinked microgels, the internal structure of the microgels is essentially independent of its presence~\cite{anderssonStructuralStudiesPoly2006}. Indeed, in all cases the experimental form factors are well-described by the fuzzy-sphere model and {\it in silico} simulations have been able to match them independently of the amount of employed surfactant~\cite{hazraStructureResponsiveMicrogels2024a,buratti2025fine}.
This suggests that the primary role of the surfactant is limited to colloidal stabilization rather than structural modification, which may be due to the fact that the reactivity ratios between BIS and NIPAM are not sufficiently far apart.
Conversely, the reactivity mismatch between EGDMA and NIPAM is more pronounced, raising the question whether SDS may actively influence the internal architecture during particle growth.  In addition, the size of the core could grow, preventing the observation of the star-like regime, even at low crosslinker concentration. Therefore, in the present work we seek to determine whether the surfactant stabilizes the growth and size of the EGDMA-rich core, and how this, in turn, influences the final particle architecture, whether star-like or more core-structured.
Next, we also address the role of crosslinker density in governing the transition 
from a pure star-like morphology to a more core-structured one. Is a minimum amount of $c$  needed and how does this depend on the amount of surfactant? To this aim, a detailed knowledge of the internal architecture as a function of crosslinker concentration is deemed necessary to be able to define the optimal conditions for achieving star-like behavior.\\
\noindent To pursue these objectives, in this work we carry out a thorough investigation combining Dynamic Light Scattering (DLS) and Small-Angle X-ray Scattering (SAXS) across the VPT, for several microgel samples, and we compare the results to monomer-resolved numerical simulations using an {\it in silico} method, recently generalized to the case of EGDMA-crosslinked microgels \cite{ballinStarLikeThermoresponsiveMicrogels2025}.
We are thus able to establish a direct correspondence between synthesis conditions and internal structure, which can be used as a guideline in the future to prepare soft thermoresponsive particles with star-like characteristics, paving the way for their use as model systems and in practical applications.

\section{Methods}
\subsection*{Microgel Synthesis}
Three sets of microgels with varying crosslinking densities were synthesized via the precipitation polymerization method: two using different amounts of the surfactant sodium dodecyl sulfate (SDS) and a third without the addition of the surfactant. For each series, the molar percentage of the crosslinker ethylene glycol dimethacrylate (EGDMA) was systematically varied.
For the synthesis, N‑isopropylacrylamide (NIPAM) monomer, the specified amount of EGDMA and possibly the surfactant SDS were dissolved in 26.5 mL of Milli‑Q water in a 50 mL two‑necked reactor fitted with a condenser and magnetic stirrer. The reactor was immersed in a thermostated oil bath. The solution was purged with nitrogen for one hour at room temperature before being heated to 70$\celsius$. Polymerization was then initiated by the controlled addition (1 mL/min) of an aqueous solution of potassium persulfate (KPS, 11.6 mg in 1.2 mL of deoxygenated water). The reaction proceeded under continuous stirring at 70$\celsius$ for five hours.
The resulting PNIPAM microgels were purified via dialysis (cellulose membrane, MWCO: 6–8 kDa) against ultrapure water for two weeks, with the water being changed twice daily. Finally, the purified microgel dispersions were freeze‑dried and stored in the dark at 4$\celsius$. 
For the two sets of samples synthesized with the surfactant, 12.73 mg (1.6 mM)  and 3.18 mg (0.4 mM) of SDS were added to the reaction mixture. Syntheses without SDS were found to yield stable microgels only up to 2\% EGDMA; for higher crosslinker concentrations, the suspension was found to contain large aggregated microgel clusters, so that these samples are not reported in the following. On the other hand the addition of even a small amount of surfactant (0.4 mM) enables stable synthesis at EGDMA concentrations of up to 10\%. The precise quantities of all reagents for each synthesis are detailed in Table \ref{tab:reagents}.

\begin{table}[!ht]
    \centering
    \setlength{\tabcolsep}{4pt} 
    \begin{tabular}{crrrc}
        \hline
        \multicolumn{2}{c}{SDS} & \multicolumn{2}{c}{EGDMA} & NIPAM  \\
        $c_\text{SDS}$  & \multicolumn{1}{c}{$m$}  & \multicolumn{1}{c}{molar}      & \multicolumn{1}{c}{$V$}      & $m$  \\
        (mM) & \multicolumn{1}{c}{(mg)} & \multicolumn{1}{c}{fraction} & \multicolumn{1}{c}{(\textmu L)} & (mg) \\
        \hline
        1.6 & 12.73 & 0.5\%$\quad$ &  4.16$\;\;$ & 497.22 \\
        1.6 & 12.73 & 0.7\%$\quad$ &  5.83$\;\;$ & 496.22 \\
        1.6 & 12.73 &   1\%$\quad$ &  8.33$\;\;$ & 494.70 \\
        1.6 & 12.73 &   2\%$\quad$ & 16.7$\;\;$ & 489.72 \\
        1.6 & 12.73 &   3\%$\quad$ & 25.0$\;\;$ & 484.72 \\
        1.6 & 12.73 &   5\%$\quad$ & 41.6$\;\;$ & 474.70 \\
        1.6 & 12.73 &   7\%$\quad$ & 58.3$\;\;$ & 464.73 \\
        1.6 & 12.73 &  10\%$\quad$ & 83.3$\;\;$ & 449.70 \\
        \hline
        0.4 &  3.18 &   1\%$\quad$ &  8.33$\;\;$ & 494.70 \\
        0.4 &  3.18 &  10\%$\quad$ & 83.3$\;\;$ & 449.70 \\
        \hline
        --  & \multicolumn{1}{c}{--} &   1\%$\quad$ &  8.33$\;\;$ & 494.70 \\
        --  & \multicolumn{1}{c}{--} &   2\%$\quad$ & 16.7$\;\;$ & 489.72 \\
        \hline
    \end{tabular}
    \caption{Reagents used in the microgel syntheses.}
    \label{tab:reagents}
\end{table}

\subsection*{Dynamic Light Scattering}
The hydrodynamic radius, $R_H$, was measured as a function of temperature using Dynamic Light Scattering (DLS). Measurements were performed with a NanoZetaSizer apparatus (Malvern Instruments LTD) equipped with a He-Ne laser (5 mW power, 633 nm wavelength) that collects light at an angle of 173$\degree$.  
Measurements were carried out by diluting the samples at a concentration of C = 0.01 wt$\%$ in MilliQ water in the temperature range between $20 \celsius$ and $45 \celsius$. After each temperature change, we equilibrated the  sample for 5 minutes.  Hydrodynamic radius $R_H$ was determined by cumulant analysis~\cite{koppel1972analysis}.
To describe the swelling behavior, we fit the hydrodynamic radius versus temperature with the function~\cite{delmonteTwostepDeswellingVolume2021a}:
\begin{equation}
    R_H(T)=R_0 - \Delta R\tanh{(s(T-T_c))}+A(T-T_c)+A_1(T-T_c)^2+A_2(T-T_c)^3
    \label{eq:RH(T)}
\end{equation}
where $R_0$ is the radius of the microgel at the VPT, $T_c$ is the VPT  temperature (VPTT), $\Delta R$ is the amplitude of the VPT and the parameter $s$ quantifies its sharpness. The function includes a third-order polynomial inserted to describe well the trend of the hydrodynamic radius over the entire temperature range, even far from $T_c$.
To characterize the VPT, we also evaluate the swelling ratio $S_R=R_H(T=20 \celsius)/R_H(T=45 \celsius)$.

\subsection*{Small Angle X-ray Scattering}
Small Angle X-ray Scattering (SAXS) was performed at the ID02 beamline~\cite{narayanan2022} of the European Synchrotron Radiation Facility (ESRF)~\cite{DatiESRF}. 
Samples were measured at low concentration of 0.1 wt\%, ensuring direct measuring of the microgel form factor.
Indeed, the SAXS scattered intensities can be expressed as:
\begin{equation}
    I(q)= \phi V(\Delta\rho)^2P(q)S(q)
\end{equation}
where $\phi$ is the particle volume fraction, V is the particle volume, $\Delta\rho$ the scattering length density difference between the microgels and the solvent, $P(q)$ the particle form factor, and $S(q)$ the structure factor, which can be considered $\approx 1$ as measurements were  performed in diluted condition.
For measurements, samples were placed in quartz capillaries (2mm diameter). 
We acquired SAXS curves using a Eiger2X 4M pixel detector and a beam energy of  12.23 keV,  at temperatures selected between $25\celsius$ and $45 \celsius$ using a Peltier stage; after each temperature change, samples were equilibrated for 5 minutes before measurements. The sample-detector distance was set to 5 m for the smaller samples synthesized with 1.6 mM of SDS and at a distance of 31 m for the bigger samples synthesized with 0.4 mM of SDS or without surfactant. The $q$-range achieved is of $1 \cdot 10^{-2} \, \text{nm}^{-1} \leq q \leq 1 \, \text{nm}^{-1}$ for the 5 m distance and of  $2 \cdot 10^{-3} \, \text{nm}^{-1} \leq q \leq 2 \cdot 10^{-1}\, \text{nm}^{-1}$   for the 31 m distance,  where $q$ is defined as $q=(4\pi/\lambda)\sin\theta$, $2\theta$ is the scattering angle, and $\lambda$ is the wavelength of the radiation. The exposure time for acquisitions was set to 0.5 s and 10 scattering patterns were acquired for each temperature.
Scattering patterns of a capillary filled with water were recorded for background subtraction.
The processing and averaging of the scattering patterns were performed by the software SAXSutilities~\cite{sztucki2021SAXSutilities2}.
Curve fitting was carried out using the software SasView~\cite{SasView}, employing different models for particle form factors, as described in detail in the Results section.

\subsection*{Numerical Methods}
{\it In silico} microgels are prepared following previous works~\cite{gnan2017silico, ninarello2019modeling}. In particular, the self-assembly of a binary mixture of $N$ attractive patchy particles of size $\sigma$ is performed in a spherical cavity of radius $Z$, such that the number density $N/(4\pi/3Z^3)\sim 0.08\sigma^{-3}$, using the oxDNA simulation package~\cite{poppleton2021oxdna}. The assembly is carried at low enough temperature to ensure the formation of a large, disordered network, including most of the particles. The mixture is composed by $cN$ crosslinking molecules with valence four, and $N(1-c)$ particles with valence two, describing NIPAM.  To model PNIPAM-BIS microgels, the crosslinkers experience a phenomenological force that pushes them towards the center of the simulation box in order to produce a rather dense core, rich in crosslinkers, to yield the typical fuzzy sphere internal structure~\cite{ninarello2019modeling}. This force was recently generalized to mimic the fast polymerization reaction of EGDMA~\cite{ballinStarLikeThermoresponsiveMicrogels2025} and validated against experiments for c=1\% and 10\% for microgels synthesized with added surfactant. Briefly, this force induces a strong confinement of the crosslinkers within a very narrow volume in the center of the microgel. The radius of this volume is set to $Z_{core}=2\sigma$ for c=1\% microgels with $N=42000$ monomers and from this value it varies as $Z_{core}(c)= Z_{core}(c=1\%) \times c^{1/3}$. 
In this work, we use the same identical force employed in Ref.~\citenum{ballinStarLikeThermoresponsiveMicrogels2025}  for additional values of $c$ and for two different microgel sizes, $N=42000$ and $N=336000$, respectively, to compare with experiments in the presence or in the absence of added surfactant. Importantly, for simulations with $N = 336000$, we use values of $Z$ and $Z_{core}$ appropriately rescaled relative to that used for $N = 42000$, in order to maintain a constant number density within each volume.
The assembly process is left to evolve until the system reaches a large number of formed bonds. For standard BIS-crosslinked microgels, we typically reach a value of formed bonds $>99.5\%$ of the total possible ones. At this point, a single network containing most of the monomers with a structure that remains rather similar at subsequent times is obtained. Afterwards, the network is frozen and the patchy interactions are replaced by the well-established bead-spring potential for polymers~\cite{grest1986}, which is complemented by a solvophobic attraction, modulated by the effective temperature $\alpha$ to take into account the worsening of the solvent with increasing temperature~\cite{soddemann2001}. The mapping between $\alpha$ and real temperature for EGDMA-crosslinked microgels is was also established in Ref.~\citenum{ballinStarLikeThermoresponsiveMicrogels2025} and is kept the same in the present work.
However, it turns out that the fast polymerization of EGDMA, accumulating at the center of the microgels, induces a very slow dependence of the microgel internal structure on the assembly time, so that the latter displays aging features at long times. We thus monitor several intermediate networks along the assembly process and select the optimal assembly time yielding agreement with experiments. We find that this time, as expected, varies with the size of the microgels, but is identical for microgels with the same total number of monomers and different crosslinker concentrations. Thus, although our computational synthesis cannot reproduce the experimental one, it still provides insights on the products of the polymerization process. In particular, we find that the final number of monomers in the simulations mimicking PNIPAM-EGDMA microgels is much smaller than the nominal one, corresponding to a value of about 70\% of the initial quantity. This is rather different from the situation observed for the modelling of PNIPAM-BIS microgels, for which more than 90\% of reagents are found to be in the final network by our standard assembly protocol.
Since all the crosslinkers react very fast, this induces a slightly higher effective value of $c$ with respect to the nominal one for the simulated EGDMA-crosslinked microgels.

\section{Results and Discussion}
\subsection{Microgels synthesized without surfactant}
\begin{figure}[H]
    \centering
    \includegraphics[width=1\linewidth]{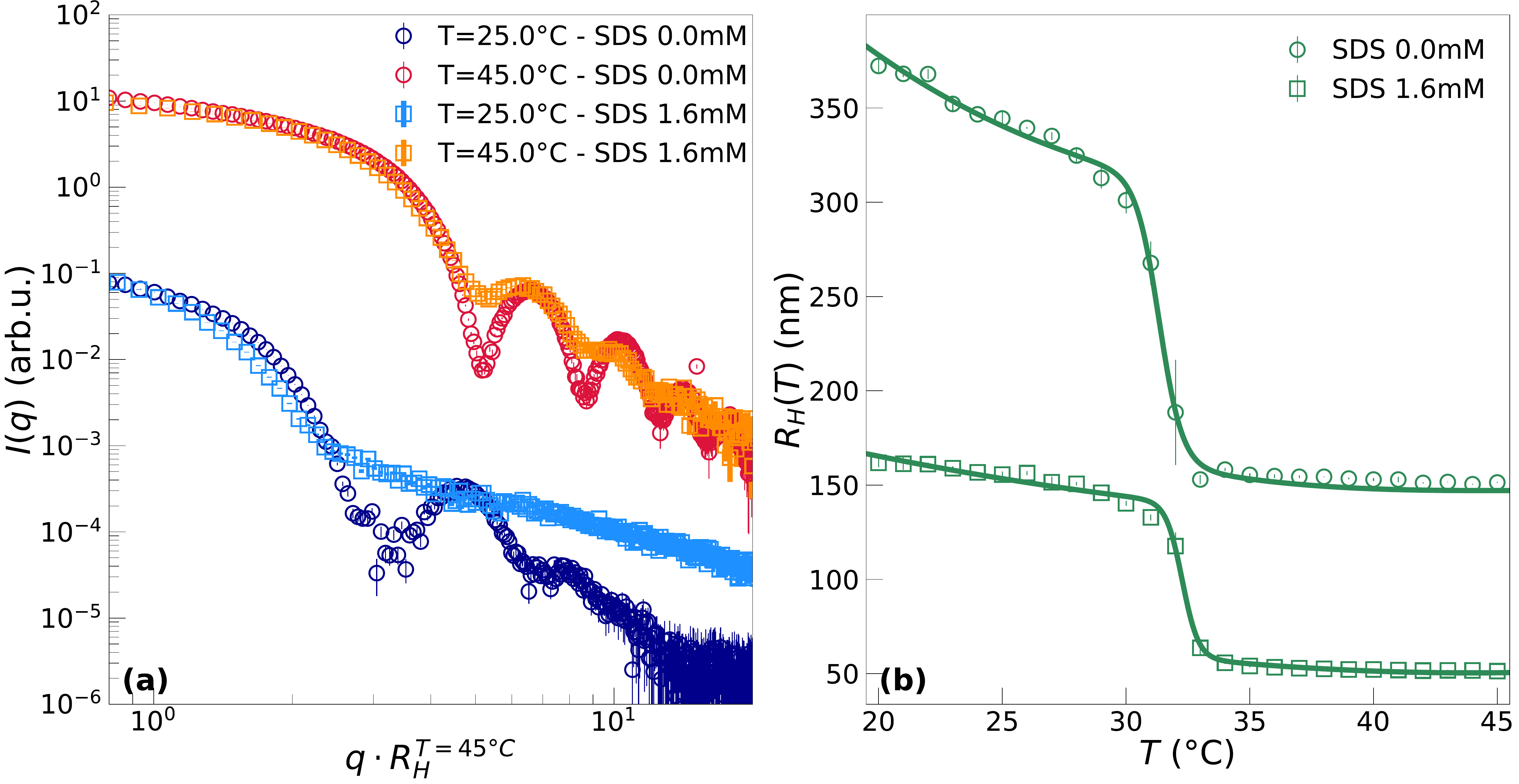}
    \caption{Characterization of EGDMA-crosslinked microgels synthesized with $c=1\%$ and with either 1.6 mM of SDS or without added surfactant. In (a) SAXS form factors at $T=25\celsius$ and $T=45\celsius$ are plotted as a function of $q\cdot R_H^{T=45\celsius}$ and shifted arbitrarily in the \textit{y}-directio; (b) Hydrodynamic radius as a function of temperature $R_H(T)$ obtained by DLS. Solid lines are fits according to Eq.~\ref{eq:RH(T)}. }
    \label{fig:EGDMA1_confrontoSDS}
\end{figure}

To investigate the role of the surfactant in the formation of star-like microgels, we first focus on samples with low crosslinker density synthesized in the absence of SDS, which are compared to microgels
obtained using 1.6 mM SDS. As put forward in the Methods section, without using surfactant  we are able to prepare stable microgels only for $c\leq2$\%.
In Ref.~\citenum{ballinStarLikeThermoresponsiveMicrogels2025} we successfully demonstrated that 
at low temperature the form factor of microgels with $c=1$\% synthesized using surfactant is fully compatible with that of a star polymer. In addition, the study of effective interactions of the \textit{in silico} star-like microgels also confirmed the Gaussian nature of their interactions, typical of star polymers, as opposed to the Hertzian-like ones of standard microgels~\cite{papettiStarlikeMicrogelsVs2026}.
However, it turns out that it is crucial to work with small enough microgels to retain the star-like structure, which is lost in the absence of surfactant. This is shown in Fig.~\ref{fig:EGDMA1_confrontoSDS} (a), where the direct comparison between SAXS curves collected on the two microgels with $c=1\%$, one with 1.6 mM of SDS and one without SDS, at $T=25\celsius$ and $T=45\celsius$ is reported. To better visualize the differences between the two samples, the $I(q)$ data are plotted as a function of $q\cdot R_H^{T=45\celsius}$, where $R_H^{T=45\celsius}$ is the respective hydrodynamic radius at $T=45\celsius$, and arbitrarily scaled on \textit{y}-axis to start from a common value at low $q$.
In this way, it is possible to immediately discriminate the size effects in the measured curves. 
At high temperature, it is evident that the peaks of $I(q)$ are found to occur at identical $q\cdot R_H$-positions, with less developed peaks occurring for the smaller microgels, due to the less resolved inner structure. 
Instead, the data at $T=25\celsius$ display marked differences in the intermediate $q\cdot  R_H^{T=45\celsius}$-range ($ 3\leq q\cdot  R_H^{T=45\celsius}\leq 10$): while the sample prepared with added surfactant shows a peak-less profile, characteristic of a star-like structure, the one synthesized without SDS exhibits a prominent peak. This signature indicates a different structure, which we hypothesize to be due to the presence of a significantly larger core, 
probably formed during the synthesis, when SDS is not present to stabilize small crosslinker-rich nuclei within the initial regime of the polymerization process.
Together with this pronounced structural difference, the swelling curves, reporting the hydrodynamic radii as a function of temperature in Fig.~\ref{fig:EGDMA1_confrontoSDS} (b), also display important variations. As expected, the presence of the surfactant strongly reduces the size of the microgels. Both curves can be fitted using Equation~\ref{eq:RH(T)}, with the extracted parameters being reported in Table~\ref{tab:parameters_rh}. In particular, the critical temperature $T_c$ is found to be slightly lower in the absence of SDS, a feature that is accompanied by a lower swelling ratio $S_R$: while in the presence of surfactant $S_R$ even exceeds 3, we find a value close to 2.5 without SDS, which is roughly compatible with the value usually reported for standard, BIS-crosslinked microgels~\cite{delmonteTwostepDeswellingVolume2021a}.
This difference is attributed to the fact that the presence of SDS stabilizes a larger number of crosslinker-rich cores, smaller in size, leaving NIPAM to form long side chains. Conversely, the incorporation of NIPAM within larger cores, which would explain the features emerging at intermediate $q\cdot R_H^{T=45\celsius}$ in the SAXS curve at 25\celsius, may cause the reduction of the arm length in the samples synthesized without SDS. These shorter chains are then responsible for the less pronounced deswelling ability of the microgels, also explaining the lower value of the sharpness transition parameter $s$ (see Table~\ref{tab:parameters_rh}).

\subsection{The need of a new model in the absence of added surfactant and comparison with numerical simulations}
\begin{figure}[H]
    \centering
    \includegraphics[width=1\linewidth]{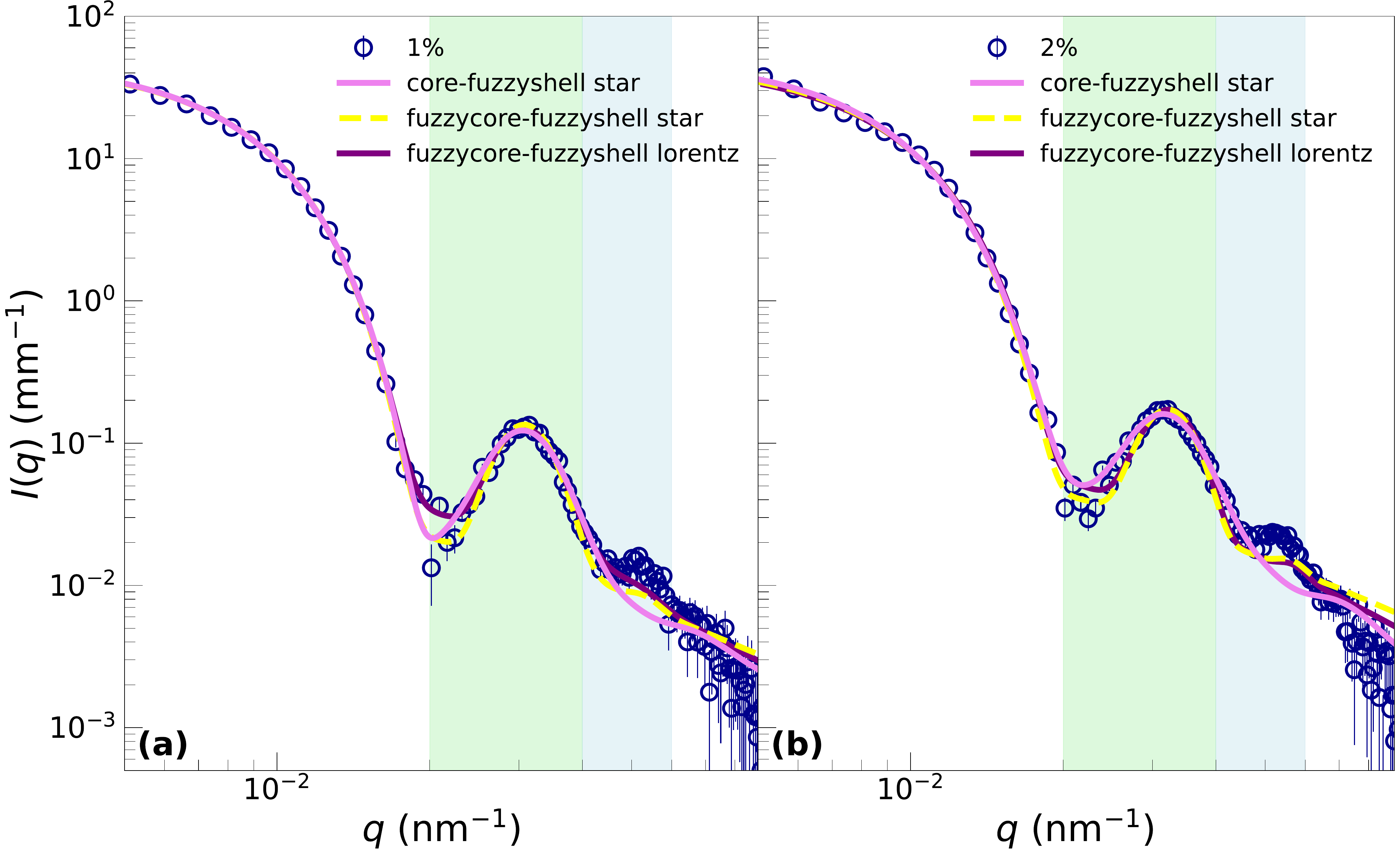}
    \caption{Comparison between experimental form factors (symbols) measured at $T=25\celsius$ for EGDMA-crosslinked microgels synthesized without SDS and the different models described in the text:  core-fuzzyshell star model (Eq.~\ref{eq:model_c-fs-star}, solid pink lines), fuzzycore-fuzzyshell star model  (Eq.~\ref{eq:model_fc-fs-star}, dashed yellow lines) and fuzzycore-fuzzyshell Lorentz  model (Eq.~\ref{eq:model_fc-fs-lorentz}, purple lines).  Data refer to microgels with $c=1\%$ in panel (a) and 2\% in panel (b). The green and blue regions highlight the peak and the shoulder, respectively,  discussed in the text.}
    \label{fig:EGDMA_confrontofitmodels_SDS0.0mM}
\end{figure}

\noindent To better investigate what happens to the microscopic structure of the samples synthesized without surfactant, we focus on the low-temperature form factors of microgels with both $c=1\%$ and 2\%, reported in Fig.~\ref{fig:EGDMA_confrontofitmodels_SDS0.0mM} (a) and (b) respectively.
\noindent A pronounced first peak appears in the $q$-range between $2 \cdot 10^{-2}$ nm$^{-1}$ and $4 \cdot 10^{-2}$ nm$^{-1}$ (highlighted by the green band), followed by a shoulder in the $q$-range between $4 \cdot 10^{-2}$ nm$^{-1}$ and $6 \cdot 10^{-2}$ nm$^{-1}$ (highlighted by the blue band).
These features were absent in the samples synthesized with surfactant (see Fig.~\ref{fig:EGDMA1_confrontoSDS}).
To model this behavior we tried different models, all reported in Fig.~\ref{fig:EGDMA_confrontofitmodels_SDS0.0mM}. 
We start by using the core-fuzzyshell star model~\cite{ballinStarLikeThermoresponsiveMicrogels2025} to fit the form factors of microgels synthesized with 1.6 mM SDS, which consists of two terms:
\begin{equation}
    P(q)= A_1P_{mgel}^{c-fs}(q) + A_2P_{star}(q).
    \label{eq:model_c-fs-star}
\end{equation}
The first term, $P^{c-fs}_{mgel}$, describes the overall particle structure, while the second term, $P_{star}$, accounts for short-range correlations within the polymer network. The two contributions are weighted by the constants $A_1$ and $A_2$. The term $P^{c-fs}_{mgel}$, where the subscript \textit{c-fs} denotes a core with a fuzzy outer shell, represents the form factor of particles modeled as a core surrounded by a fuzzy shell, and reads as, 
\begin{equation}
    P^{c-fs}_{mgel}(q)= \frac{1}{V_{tot}}\,\left [ (\rho_c-\rho_s)A_s(q,r_c)+(\rho_s-\rho_0)A_s(q,r_s)\exp\left(-\frac{1}{2}\sigma_s^2q^2\right) \right ]^2\quad.
    \label{eq:core-fuzzy-shell_model}
\end{equation}
Here $r_c$ is the core radius with scattering length density $\rho_c$; $r_s$ is the outer radius of the fuzzy shell, characterized by a fuzziness parameter $\sigma_s$ and scattering length density $\rho_s$; $\rho_0$ is the scattering length density of the solvent; and $V_{\text{tot}} = \frac{4}{3}\pi R_t^3$ is the total volume of the particle, with $R_t \approx r_s + 2\sigma_s$~\cite{stiegerSmallangleNeutronScattering2004}.  $A_s(q,R)$ denotes the scattering amplitude of a homogeneous sphere of radius $R$ and is given by:
\begin{equation}
A_s(q,R) = 3V \frac{\sin(qR)-qR \cos(qR)}{(qR)^3}.
\end{equation}
In this model the thickness of the shell is $t=r_s-r_c$.
The second term of Eq.~\ref{eq:model_c-fs-star}, $P_{star}(q)$, is the form factor proposed by Dozier et al.~\cite{dozierColloidalNatureStar1991} for star polymers:
\begin{equation}
    P_{star}(q) = \frac{4\pi}{q\xi} \frac{\sin[\mu \arctan(q\xi)]}{(1+q^2\xi^2)^{\frac{\mu}{2}}} \Gamma\left(\frac{\mu}{2}\right) \quad 
    \label{eq:star_model}
\end{equation}
where $\mu = (1 / \nu) - 1$, with $\nu$ the Flory solvency parameter and $\xi$ the maximum size of the polymer chain blobs.
We find that Eq.~\ref{eq:model_c-fs-star} is able to describe the first peak (see Fig.~\ref{fig:EGDMA_confrontofitmodels_SDS0.0mM}), but fails to capture the additional shoulder, highlighted in the light blue region of  Fig.~\ref{fig:EGDMA_confrontofitmodels_SDS0.0mM}. Indeed, while the model can produce a small shoulder, its position is constrained by the first peak location and, therefore, it cannot be adjusted to match the measured position of the additional peak. 
Therefore, to adequately capture the experimental form factor we resort to a different model. We hypothesize that the absence of surfactant removes a structural constraint, allowing for a more articulated internal architecture. To model this, we introduce a fuzzy interface between the core and the shell, which amounts to generalizing Eq.~\ref{eq:core-fuzzy-shell_model} by replacing $P_{mgel}^{c-fs}(q)$ with the new expression $P_{mgel}^{fc-fs}(q)$, 
\begin{equation}
    P_{mgel}^{fc-fs}(q)= \frac{1}{V_{tot}}\,\left [ (\rho_c-\rho_s)A_s(q,r_c)\exp\left(-\frac{1}{2}\sigma_c^2q^2\right) +(\rho_s-\rho_0)A_s(q,r_s)\exp\left(-\frac{1}{2}\sigma_s^2q^2\right) \right ]^2\quad,
    \label{eq:fuzzycore-fuzzyshell_model}
\end{equation}
in which the superscript \textit{fc-fs} is an abbreviation for fuzzycore-fuzzyshell. This term thus describes the form factor of core-shell particles in which both the core and the shell have a fuzzy interface. Clearly, this new expression  differs from the previous one  only for the addition of the fit parameter  $\sigma_c$, which represents the fuzziness of the core, thus including the previous model as a more general case.
The  fuzzycore-fuzzyshell star model is then calculated as the sum of $P_{mgel}^{fc-fs}(q)$  and $P_{star}(q)$:
\begin{equation}
    P(q)= A_1P_{mgel}^{fc-fs}(q) + A_2P_{star}(q).
    \label{eq:model_fc-fs-star}
\end{equation}
The fit of experimental data to this model, also reported in Fig.~\ref{fig:EGDMA_confrontofitmodels_SDS0.0mM}, allows for an independent modulation of the shoulder position with respect to the main peak, improving the agreement with experimental data.
For completeness, we also reconsider the $P_{star}(q)$ term to investigate whether it is crucial or it can be replaced by a more standard Lorentzian term, normally used to describe the microscopic structure of the polymer network of standard microgels. We thus also report results for the so-called fuzzycore-fuzzyshell Lorentz model:
\begin{equation}
    P(q)= A_1P^{fc-fs}_{mgel}(q)+\frac{A_2}{1+(q\Xi)^2},
    \label{eq:model_fc-fs-lorentz}
\end{equation}
where $\Xi$ is the average correlation length, that provides an estimate of the average mesh of the network. We basically show in Fig.~\ref{fig:EGDMA_confrontofitmodels_SDS0.0mM} that there is no difference substituting the star term with the Lorentzian one. Thus, both models can be used to describe the experimental data. We also note that a similar theoretical description has been proposed for polymer micelles by Pedersen and Gerstenberg\cite{pedersenScatteringFormFactor1996}; however, the parameterization used therein is not directly tailored to the typical structural features of microgels, motivating us to develop the present formulation. Overall, these results indicate that the internal structure of microgels synthesized without using surfactant is markedly different from that of a star-like particle.
\begin{figure}[H]
    \centering
    \includegraphics[width=0.9\linewidth] {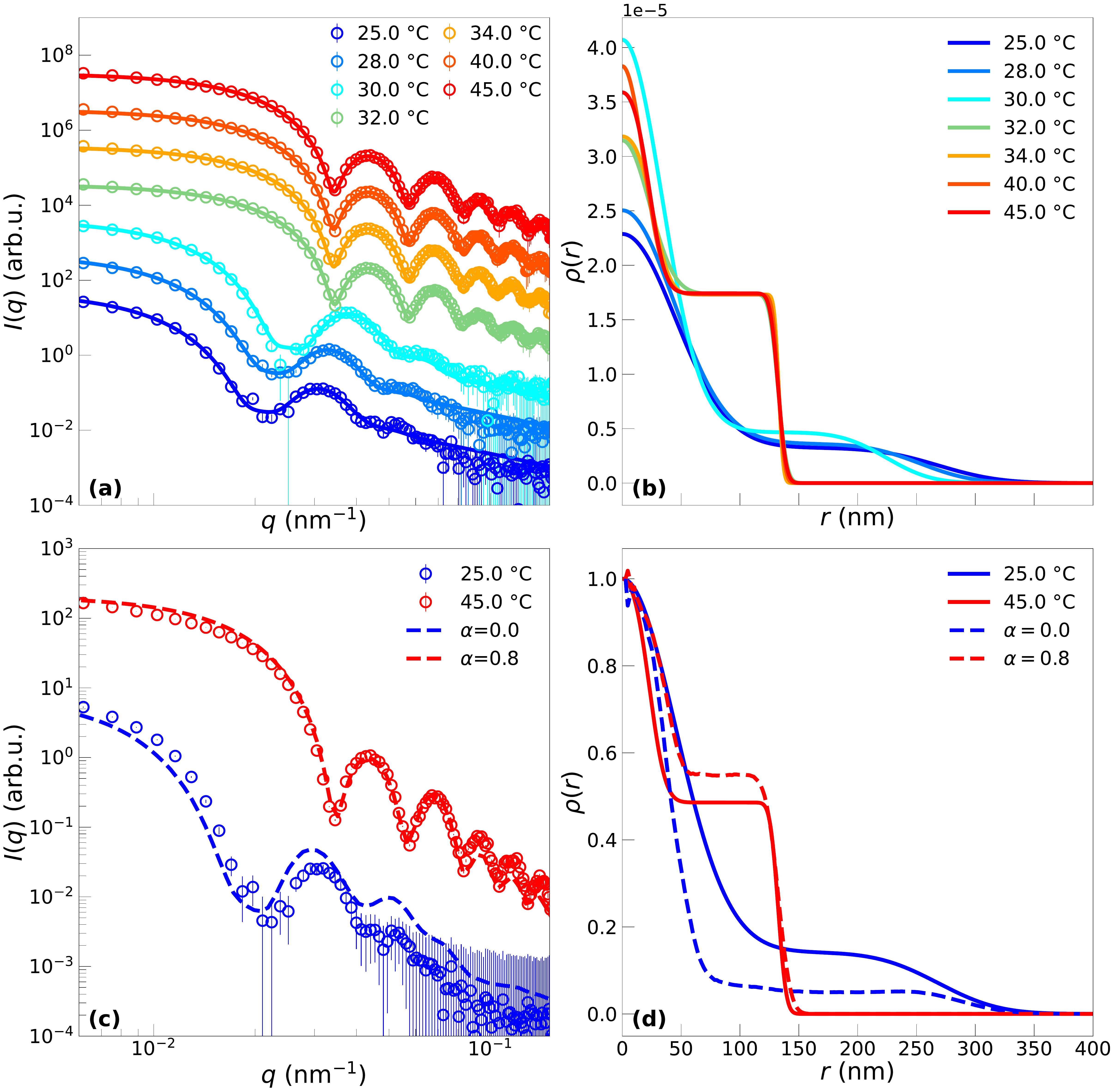}
    \caption{(a) SAXS data (symbols) for EDGMA-crosslinked microgels with $c=1\%$ synthesized without SDS across different temperatures and corresponding fits (solife lines) the fuzzycore-fuzzyshell model reported in Eq.~\ref{eq:model_fc-fs-lorentz}. To improve visualization, the curves were arbitrarily offset in the \textit{y}-direction. The unshifted data, presented in actual units, are displayed in Fig.~\ref{fig:DatiSAXS_SDS0.0mM}(a) of the SI; (b) Radial density profiles, $\rho(r)$, obtained via Eq.~\ref{eq:rho} using the best-fit parameters of the $P(q)$ shown in panel (a).  Curves are normalized so that the spherical integral of the $\rho(r)$ equals one; (c) Direct comparison of the experimental form factors (simbols) with those calculated from \textit{in silico} microgels (dashed lines) with the same nominal crosslinker concentration;  (d) Direct comparison between the experimentally derived $\rho(r)$ profiles (solid lines) and those from the simulations (dashed lines). Curves are normalized so that $\rho(0)=1$. In both panels (c) and (d), simulation data are rescaled onto experimental ones using the monomer size $\sigma = 3.08$ nm. }
    \label{fig:EGDMA1_fit_sim_rho_SDS0.0mM}
\end{figure}
\noindent The complete set of form factors of microgels with $c=1$\% at all investigated temperatures, ranging from $25 \celsius$ to $45 \celsius$, is thus reported in Fig.~\ref{fig:EGDMA1_fit_sim_rho_SDS0.0mM}(a) with the corresponding fits by Eq.~\ref{eq:model_fc-fs-lorentz}. The model is able to capture quite well the temperature evolution of the data, which 
display a rather sharp transition to a collapsed state between 30 and 32$\celsius$. 
This is confirmed by looking at the density profiles $\rho(r)$, calculated by Fourier transforming the square root of $P_{mgel}^{cf-sf}$ as,
\begin{equation}
    \rho(r)^{fc-fs}=\frac{1}{2\pi^2}\int_0^\infty \sqrt{P^{fc-fs}_{mgel}(q)}\frac{\sin(qr)}{qr}q^2 dq.
    \label{eq:rho}
\end{equation}
These are reported in Fig.~\ref{fig:EGDMA1_fit_sim_rho_SDS0.0mM} (b) for all temperatures, showing the presence of an inner core, which shrinks slightly upon increasing $T$, suggesting that it may be composed not only of crosslinkers, but also of NIPAM monomers experiencing the worsening of solvent quality. In parallel, the rest of the microgel, made of essentially long chains, also compacts giving rise to a characteristic shoulder upon crossing the VPT. These features
reinforce the idea that the core acquires a significant size in the absence of surfactant also at very low EGDMA concentration. We confirm this microscopic picture with numerical simulations of a microgel with $N=336000$ beads, assembled as described in Methods. \\

\noindent The corresponding results for $c=1\%$ as in experiments are reported in Fig.~\ref{fig:EGDMA1_fit_sim_rho_SDS0.0mM}. In particular, the calculated form factors and density profiles are shown in panels (c) and (d), respectively, for two values of the solvophobic parameter: $\alpha = 0.0$ and $0.8$. These values correspond to temperatures of $T = 25\celsius$ and $T = 45\celsius$, respectively, following the mapping already adopted in Ref.~\citenum{ballinStarLikeThermoresponsiveMicrogels2025}. The numerical data are then directly compared to the experimental ones using a monomer size of $\sigma = 3.08$ nm.
Notably, the simulations are able to successfully reproduce the features in the intermediate $q$-range observed at low temperatures, that were instead not present for smaller microgels ($N=42000$), synthesized in an identical way to the present one except for the larger number of monomers. This is an important result, as it would be unpleasant that the {\it in silico} model should be adjusted depending on the synthesis conditions. Indeed, previous work for BIS-crosslinked microgels~\cite{ninarello2019modeling} had shown no systematic dependence of the form factors on the microgel size, although the peaks become less resolved with a smaller number of monomers. Instead, size-dependence strongly manifests in the experimental form factors and is also captured by the {\it in silico} assembly, suggesting that it is a genuine feature of EGDMA-crosslinked microgels, in contrast to BIS-crosslinked ones.
This can be understood by the fact that a large enough compact core must be present within the microgels in order to observe the experimental features arising without surfactant: a large number of monomers is thus needed in order to populate the core in a large enough volume, likely also involving NIPAM monomers. In contrast, in the presence of SDS, the core remains limited in size and is therefore essentially composed of crosslinkers only. These features are also reflected in the density profiles, reported in
 Fig.~\ref{fig:EGDMA1_fit_sim_rho_SDS0.0mM} (d), which compare the numerical results with those extracted from the fits of the experimental form factors. Both models indicate the presence of the inner core, surrounded by a very large corona at low temperatures. As temperature increases, the contrast between the core and corona diminishes due to the densification and collapse of the corona. Concurrently, the corona thickness contracts, progressively erasing the fuzzy interfaces in the shell and between the core and corona. However, simulations predict a smaller shrinking of the inner core in comparison to experiments, probably due to the coarse-graining. Indeed, in reality, the number of monomers in the samples is of course much larger than in simulations. Finally, it is worth noting that the $\rho(r)$ profiles being compared are not strictly equivalent, since the experimental profile represents the scattering length density, which also depends on the specific chemical species present. Nonetheless, the agreement between experiments and simulations is qualitatively very good, thus confirming the ability of the chosen fuzzycore-fuzzysphere model to describe the experimental curves. Similar results are also found for the microgels synthesized with c=2\%. The unshifted data in real units are displayed in Fig.~\ref{fig:DatiSAXS_SDS0.0mM}(b), while Fig.~\ref{fig:EGDMA2_fit_sim_rho_SDS0.0mM} reports the form factors alongside the simulation results and the corresponding density profiles, where the simulations are performed \textit{in silico} microgels with the same nominal crosslinker concentration as in experiment. For completeness, Figure~\ref{fig:fuzzycore_fuzzyshell_parameters_SDS0.0mM} also shows the best-fit parameters for the form factor of microgels synthesized without SDS for both investigated $c$-values. Their behavior as a function of temperature is similar for both samples.

\begin{figure}[H]
    \centering
    \includegraphics[width=0.5\linewidth]{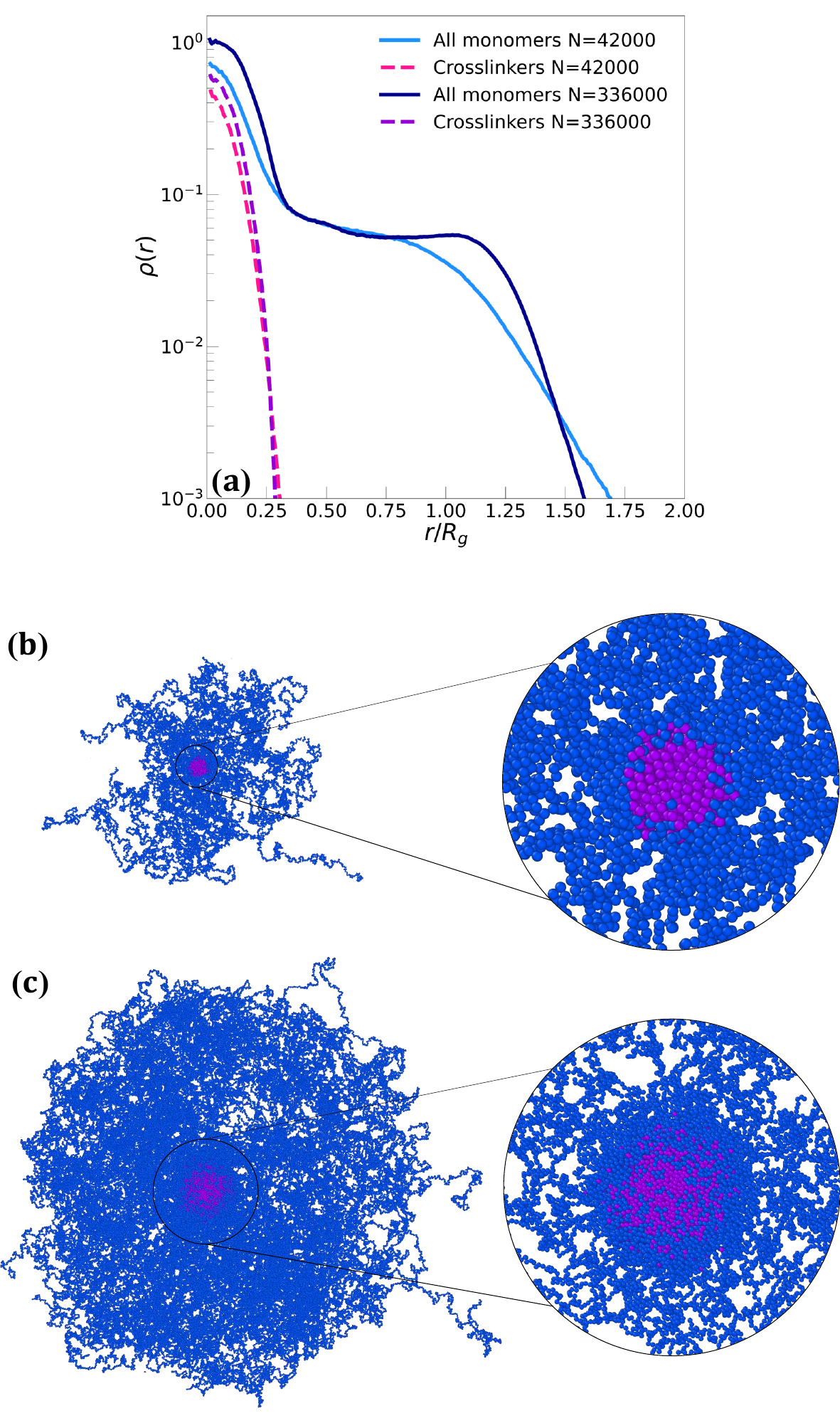}
    \caption{(a) Density profiles for the \textit{in silico} microgels with $c = 1\%$ and with either $N=42000$ or $N=336000$. Solid lines indicate profiles of all the monomers, whereas dashed lines refer to crosslinkers alone; (b) and (c) snapshots of the simulated microgels with $N=42000$ and $N=336000$, respectively. To visualize of the core region, half of each microgel is reported with uniform bead size across both systems. To highlight the core conformation, the central regions are also zoomed-in. In these cases, the bead size is not preserved and differs between the two systems.}
    \label{fig:confronto_EGDMA_e_snapshot}
\end{figure}

In order to directly compare the differences between simulations of different sizes, Figure~\ref{fig:confronto_EGDMA_e_snapshot}(a) displays the numerical density profiles of both monomers and crosslinkers for simulated microgel with with $c = 1\%$, comparing small ($N=42000$) and large ($N=336000$) microgels. To this end, the density profiles are rescaled along the 
$x$-axis by the radius of gyration $R_g$ of each microgel. The characteristic shape of the density profile of star-like objects, consisting of a power-law decay after the core followed by a Gaussian tail~\cite{ruiz-francoMultiparticleCollisionDynamics2019}, 
is observed only for the smaller microgels, whereas for the larger ones it is completely absent and the tail of $\rho(r)$ decays exponentially. In addition, there is a much sharper interface between core and corona in large microgels, which display a very flat region for intermediate values of $r$, whereas for the small system, the density profile decays almost continuously in the corona region. We further observe that a larger difference between the monomers and the crosslinkers profiles is present at short distances for large microgels, signaling the presence of many monomers also in that region. 
These features are confirmed by the snapshots shown in Fig.~\ref{fig:confronto_EGDMA_e_snapshot}(b) and (c). In particular, the smaller microgel of panel (b) exhibits a much looser external structure, characterized by long dangling chains and a rather dilute outer shell. In this case, the crosslinkers are all concentrated in the inner core, which essentially does not contain additional monomers. The microgel also displays a rather anisotropic morphology. In contrast, the larger microgel of panel (c) possesses a much denser shell and a much more spherical shape. In addition, the core region of the larger microgels is rather extended, containing  a large number of NIPAM monomers, as clearly visible in the zoom of the core region of the snapshots. Altogether these results indicate that the synthesis of EGDMA-crosslinked microgels without surfactant leads to the presence of a significant core at the center of the microgel, which leads to the loss of the star-like architecture, even at very low crosslinker concentration.

\subsection{Increasing surfactant amount to recover the star-like structure}
Having established the architecture of microgels synthesized without using surfactant, we now investigate whether the addition of a small amount of SDS can restore the star-like structure that is obtained at low crosslinker content ($c=1$\%) and high surfactant concentration (1.6 mM). To this end, we synthesized EGDMA-crosslinked microgels using 0.4 mM SDS and two different molar fractions of crosslinker, $c=1$\% and $c=10$\%. Importantly, under these conditions, we do not detect microgel aggregation also at high EGDMA content ($c=10$\%), confirming that this reduced concentration of surfactant is sufficient to stabilize the dispersions. 
\begin{figure}[H]
    \centering
    \includegraphics[width=1\linewidth]{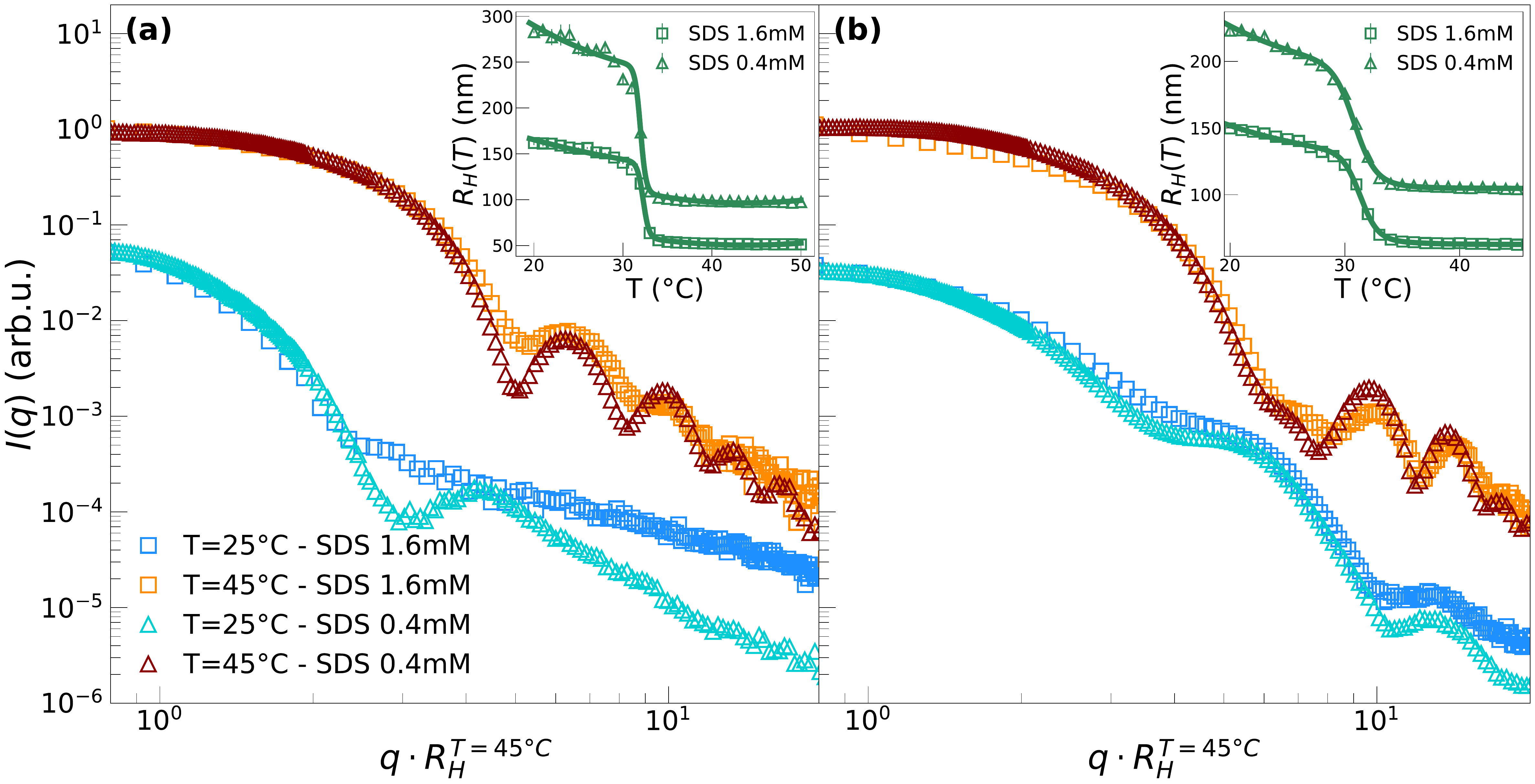}
    \caption{SAXS scattered intensities $I(q)$ measured at $T = 25\celsius$ and $T = 45\celsius$ for EGDMA-crosslinked microgels synthesized with SDS 1.6 mM and SDS 0.4 mM, at $c = 1\%$ (a) and $10\%$ (b). Each curve is rescaled along the $x$-axis by the respective value of the hydrodynamic radius at $T = 45\celsius$. For visual clarity, the curves are also arbitrarily shifted along the $y$-axis. The insets in (a) and (b) display the hydrodynamic radii as a function of temperature for the samples with $c = 1\%$ (a) and $c = 10\%$ (b). Solid lines represent fits to Eq.~\ref{eq:RH(T)}.}
    \label{fig:EGDMA1-10_confrontoSDS_temp}
\end{figure}

\begin{figure}[H]
    \centering
    \includegraphics[width=0.8\linewidth]{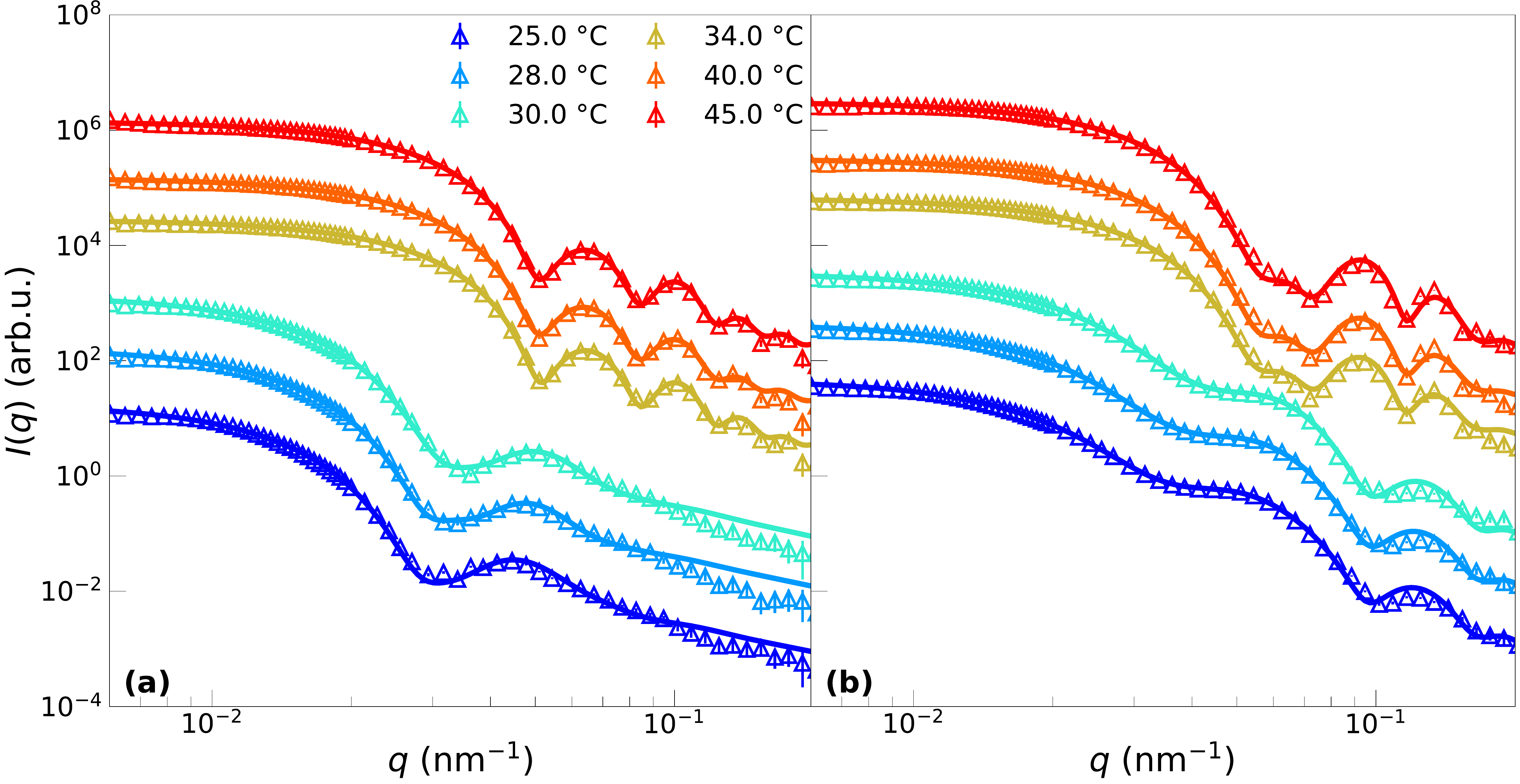}
    \caption{SAXS scattered intensities $I(q)$ (symbols) for EGDMA microgels synthesized with  SDS 0.4 mM for $c = 1\%$ (a) and $10\%$ (b) in the temperature range between  $25\celsius$ and $45\celsius$. For visual clarity each curve is shifted arbitrarily along the $y$-axis. In both panels, solid lines are the corresponding fits performed with the core-fuzzy shell model (Eq.~\ref{eq:model_fc-fs-lorentz}).}
    \label{fig:EGDMA1-10_fit_SDS0.4mM}
\end{figure}

\noindent Figure~\ref{fig:EGDMA1-10_confrontoSDS_temp} compares the form factors of the microgels synthesized with 1.6 mM and 0.4 mM surfactant, at $c = 1\%$ (a) and $10\%$ (b), measured at $T = 25\celsius$ and $T = 45\celsius$. To aid comparison, each form factor has been normalized to its respective hydrodynamic radius at $T = 45\celsius$, $R_H^{(T=45\celsius)}$.
The raw data for the samples synthesized with 0.4 mM SDS are shown in Figure~\ref{fig:DatiSAXS_SDS0.4mM}. At $c = 10\%$, the close overlap of the form factors of the two samples indicates that the internal structure is largely unaffected by the concentration of SDS in the synthesis. At high temperature, the more pronounced peaks for the sample with low SDS content reflect improved resolution of the internal structure for larger microgels.
In contrast, at $c = 1\%$, some differences emerge, particularly at low temperature. The peak-less scattering profile observed for 1.6 mM SDS becomes a broad peak centered at $qR_H^{(T=45\celsius)}\approx 4$ for 0.4 mM SDS, signaling the emergence of a more distinct core. Despite this, the core–fuzzyshell star model (Equation \ref{eq:model_c-fs-star}) is still able to provide an excellent fit across the entire temperature range (25–45\celsius) for both $c=1\%$ and $c=10\%$ (Figure~\ref{fig:EGDMA1-10_fit_SDS0.4mM} (a) and \ref{fig:EGDMA1-10_fit_SDS0.4mM} (b), respectively), in analogy to the case of 1.6mM SDS. The corresponding fit parameters are summarized in Figure~\ref{fig:star_fuzzysphere_parameters_SDS0.4mM}. 

\noindent The swelling curves for these microgels in comparison to the ones at higher SDS content are shown in the insets of Fig.~\ref{fig:EGDMA1-10_confrontoSDS_temp}(a) and (b)), for $c = 1\%$ and  $c = 10\%$, respectively. We observe that the swelling behavior of the sample containing 0.4 mM SDS closely mirrors that of the 1.6 mM SDS sample. For  $c = 1\%$, the fit parameters, reported in Table~\ref{tab:parameters_rh}, are substantially closer to those of the high-surfactant sample than to the surfactant-free one. This indicates that even a small amount of SDS is sufficient to restore the sharp swelling characteristics typically associated with the star-like regime. It is also worth noting that simulations of star polymers ~\cite{papettiStarlikeMicrogelsVs2026} similarly predict a small but finite peak in the form factor, reflecting the presence of a tiny compact core.
\begin{table}[H]
    \centering
    \begin{tabular}{|rccll|}
        \hline
        \multicolumn{1}{|c}{$c$} & $c_\text{SDS}$ & $T_c$      & \multicolumn{1}{c}{$s$}               & \multicolumn{1}{c|}{$S_R$} \\ 
                                & (mM)           & (\celsius) & \multicolumn{1}{c}{(\celsius$^{-1}$)} &                           \\ 
        \hline
        1\%   & 0.0 & $31.3\pm0.1$ & $1.1\;\:\pm0.2$ & 2.46 \\
        1\%   & 0.4 & $32.0\pm0.1$ & $1.6\;\:\pm0.2$ & 2.89 \\
        1\%   & 1.6 & $32.3\pm0.1$ & $1.5\;\:\pm0.2$ & 3.15 \\
        10\%  & 0.4 & $30.8\pm0.1$ & $0.55\pm0.05$ & 2.14 \\
        10\%  & 1.6 & $31.3\pm0.1$ & $0.65\pm0.05$ & 2.40 \\ \hline
    \end{tabular}
    \caption{Fit parameters obtained by fitting the swelling curves reported in Figures~\ref{fig:EGDMA1_confrontoSDS} (b) and in the insets of Figure~\ref{fig:EGDMA1-10_confrontoSDS_temp}(a) and (b) to Equation \ref{eq:RH(T)}. Here, $T_c$ denotes the transition temperature, $s$ quantifies the sharpness of the transition, and $S_R$ is the swelling ratio between the low- and high-temperature radii.}
    \label{tab:parameters_rh}
\end{table}

\noindent Taken together, these results demonstrate that a star-like behavior is successfully recovered at low EGDMA content, despite the core not being strictly point-like, in the presence of a small amount of SDS. Conversely, the absence of SDS causes the microgels to acquire a much more pronounced structure, more similar to a core-shell structure, losing the peculiar characteristics of star-like microgels, both in terms of internal structure, that needs to be described by a different model, and of swelling properties.
Furthermore, the addition of a small amount of SDS enables the synthesis of stable microgels  at high $c$, yielding samples whose structural features, aside from the expected difference in overall size, closely resemble those obtained with a higher surfactant content.  These findings enable the preparation of stable EGDMA-crosslinked microgels with radii around 300 nm and below. Note that, however, the absence of surfactant did not produce much larger microgels, due to their change in the corona properties, best visible in the simulation results of Fig.~\ref{fig:confronto_EGDMA_e_snapshot}.

\subsection{Varying crosslinker density at high surfactant concentration}
\begin{figure}[H]
    \centering
    \includegraphics[width=0.8\linewidth]{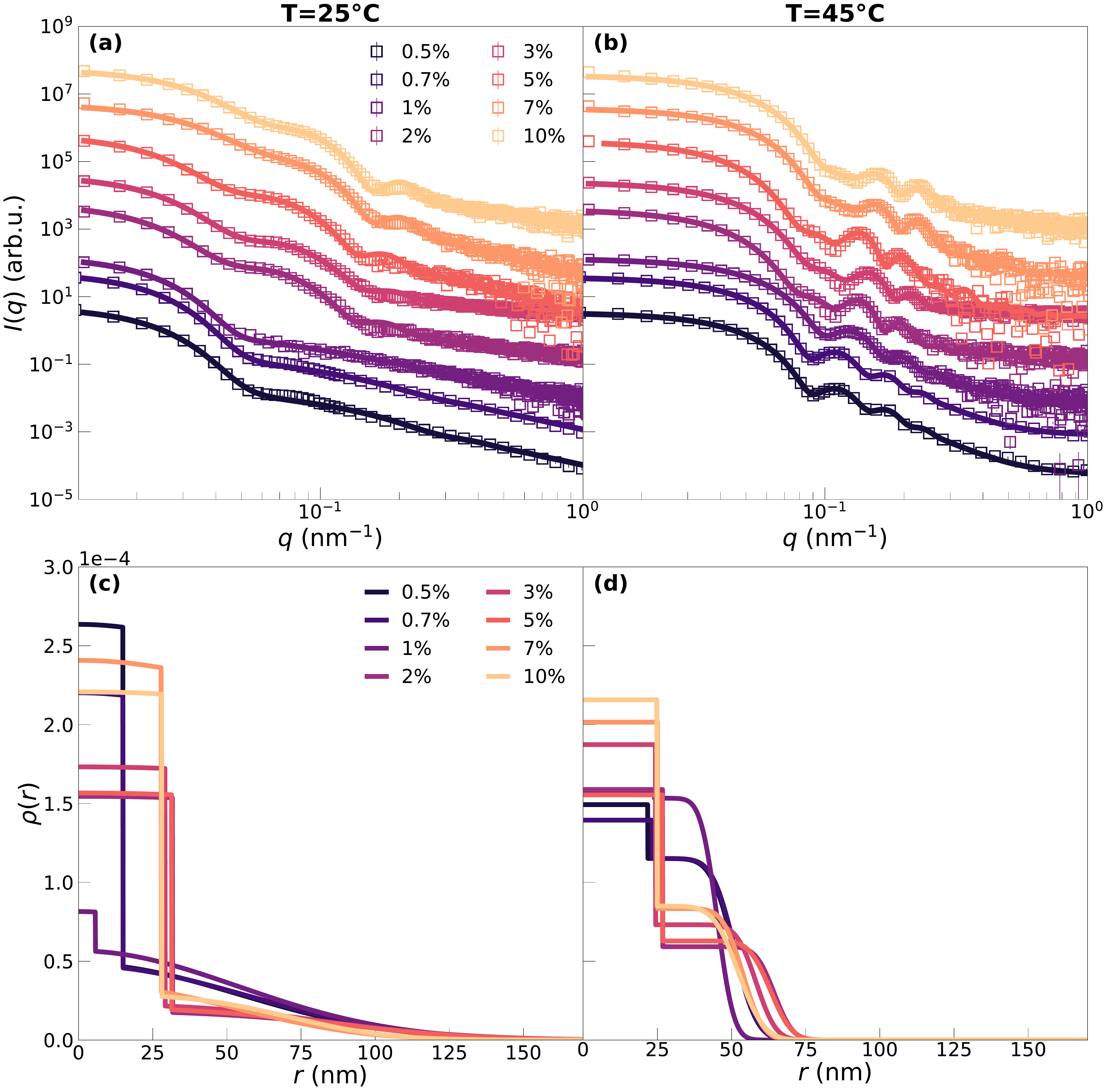}
    \caption{SAXS scattered intensities $I(q)$ (symbols) for EGDMA-cosslinked microgels synthesized with SDS 1.6 mM with $c$ between 0.5\% and 10\% at $T=25\celsius$ (a) and $T=45\celsius$ (b). Solid lines represent fits in which the form factor $P(q)$ is described by the star-like fuzzy-sphere model reported in equation \ref{eq:model_c-fs-star}. For visual clarity curves are shifted arbitrarily in the $y$ direction.  (c) and (d) radial density curves calculated as the Fourier transform of the square root of equation \ref{eq:core-fuzzy-shell_model} inserting the parameters of the fits reported in panel (a) and (b), respectively. Curves are normalized such that the spherical integral of the $\rho(r)$ equals one.}
    \label{fig:EGDMA_T25-45_fit_SDS1.6mM_normintergal}
\end{figure}

In this section we focus on microgels synthesized using high surfactant concentration (1.6 mM SDS) and varying the crosslinker density between $c=0.5\%$ and $c=10\%$, in order to characterize in detail the structural changes occurring as a function of EGDMA concentration. 
Figure~\ref{fig:EGDMA_T25-45_fit_SDS1.6mM_normintergal} reports the SAXS characterization of the microgels, with
panels (a) and (b) displaying the scattering intensity, $I(q)$, at $T = 25~\celsius$ and $T = 45~\celsius$, respectively.  The full set of measured curves at all intermediate temperatures is provided in the SI (Fig.~\ref{fig:all_fit_SDS1.6mM}(a-h)). All SAXS profiles, at both 25 and 45\celsius, can be described using the core-fuzzyshell star model of Eq.~\ref{eq:model_c-fs-star}.
The scattering curves at $25~\celsius$ reveal a structural evolution with increasing crosslinker concentration. For $c \leq 1\%$, the intensity curves lack a distinct shoulder in the intermediate $q$-range, resembling the star-like structure, with a small core and an extended corona. In contrast, for $c \geq 2\%$, a more evident shoulder emerges, indicative of the presence of a larger core. 
The distinction between the two sets of samples, $c\leq 1\%$ and $c\geq 2\%$, is preserved also at $45~\celsius$. Samples with $c \leq 1\%$ collapse into a compact, sphere-like structure, evidenced by the characteristic peaks of a sphere in their form factors. Conversely, samples with $c \geq 2\%$ exhibit an additional shoulder preceding the first peak, suggesting that a more complex internal architecture is present also in the collapsed state.

\begin{figure}[H]
    \centering
    \includegraphics[width=0.6\linewidth]{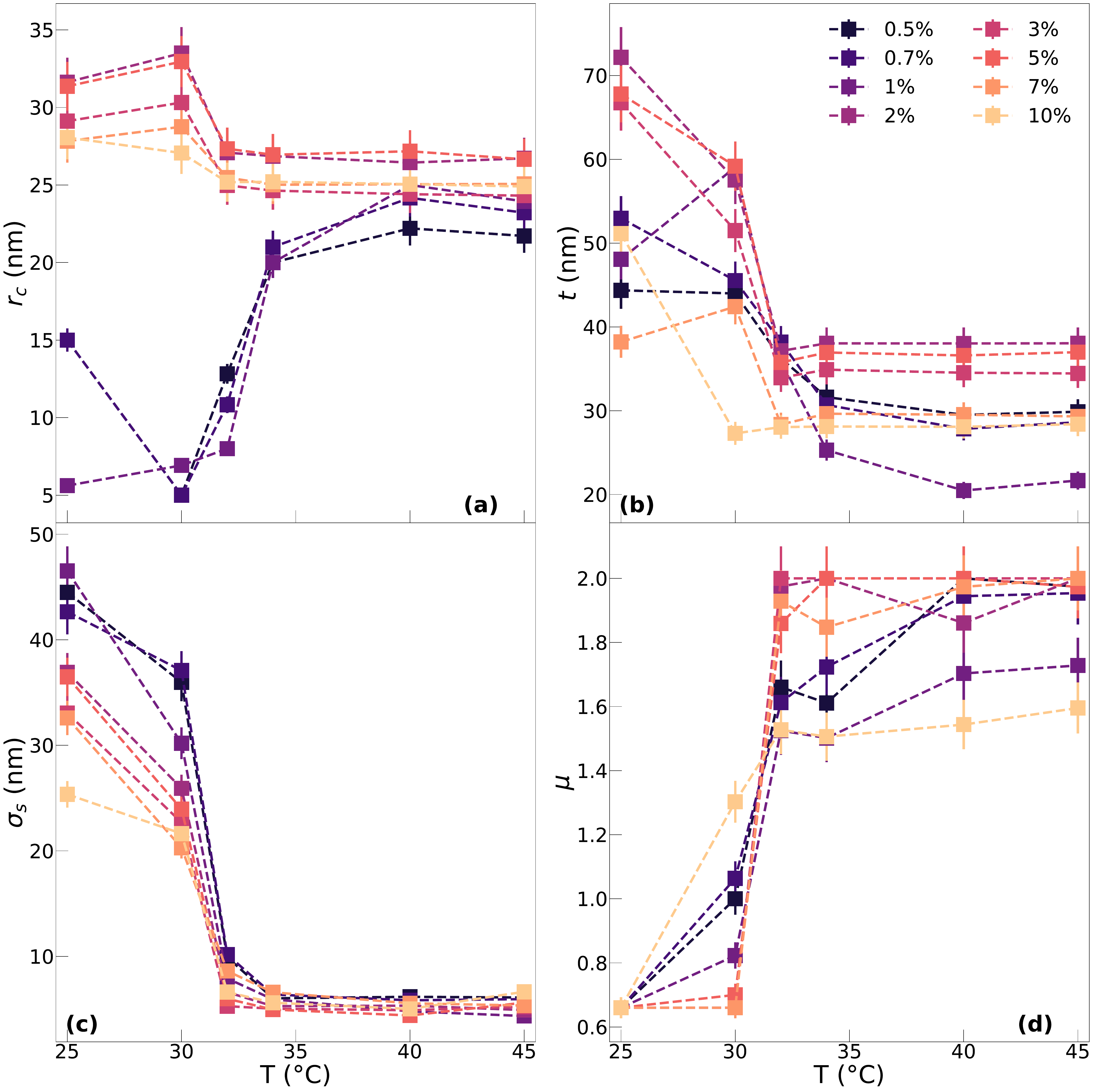}
    \caption{Fit parameters of the scattering intensity curves as a function of temperature: (a) radius of the core $r_c$; (b) thickness of the shell $t=r_s-r_c$; (c) fuzziness of the shell $\sigma_s$; (d) $\mu=(1/\nu)-1$, with $\nu $ the Flory solvency parameter. }
    \label{fig:star_fuzzysphere_parameters_SDS1.6mM}
\end{figure}

\noindent The radial density profiles $\rho(r)$ corresponding to the form factors of panels (a) and (b) of Fig.~\ref{fig:EGDMA_T25-45_fit_SDS1.6mM_normintergal} are reported in panels (c) and (d), respectively. The profiles are calculated through Eq.~\ref{eq:rho}, where $P^{fc-fs}_{mgel}(q)$ is replaced by $P^{c-fs}_{mgel}(q)$ using the best-fit parameters. The curves show that by increasing EGDMA content there is an increase of the core size, accompanied by a decrease of its density. Moreover, upon collapse, samples with $c\geq 2\%$ exhibit a well-defined core–shell architecture. In contrast, profiles for samples with $c\leq 1\%$ exhibit a significantly reduced density gap between the core and shell regions, resulting in a density profile that closely resembles that of a fuzzy sphere. We notice that this is most evident for $c =1\%$, rather than for 0.7\% and 0.5\% which display a slightly more evident peak in the form factor and a slightly larger difference between core and corona structure in $\rho(r)$. This suggests that the fast polymerization of EGDMA can lead to variations in the synthesized microgels due to the different accumulation in the core region. 
For a more quantitative description, we analyze the temperature dependence of the best-fit parameters obtained from fitting the $I(q)$ form factors. These are shown in Figure \ref{fig:star_fuzzysphere_parameters_SDS1.6mM}, providing information on the structural evolution of the microgels with increasing $c$ and with temperature.
Panel (a) shows the trend of the core radius, $r_c$. For microgels with $c \leq 1\%$, $r_c$ is initially small ($\leq 15$ nm) and increases as the temperature increases above the VPTT. This occurs because, at low temperature, these particles possess a very tiny core. Upon heating above the VPTT, the entire microgel collapses into a nearly spherical particle, leading to an increase in the apparent core size. Since the density difference between the core and the shell is very small, as previously noted in the discussion of Figure \ref{fig:EGDMA_T25-45_fit_SDS1.6mM_normintergal}(d), there is little distinction between the core and shell regions in the collapsed state. In contrast, for samples with $c\geq 2\%$, $r_c$ starts at a value that is roughly twice that of the low crosslinker microgels at low temperature and remains almost constant upon heating. This indicates that these microgels possess a more defined, structured core. During collapse, this core retains its distinctness and is largely unaffected by the collapse of the outer shell.
Notably, within the low crosslinker regime, $r_c$ does not increase coherently with the nominal $c$. The scattering curve for $c=1\%$ is essentially featureless in the intermediate $q$-range, allowing it to be fitted with a wide range of small $r_c$ values. Conversely, the curves for $c=0.5\%$ and $0.7\%$ show a very faint shoulder, which constrains the fit to a finite, albeit small, core size. This non-monotonic trend and the appearance of an extremely weak shoulder may result from variations in effective crosslinker incorporation during synthesis.
Panel (b) displays the temperature dependence of the corona thickness, $t$, which generally decreases with $T$ for all samples. Next, panel (c) shows the evolution of the parameter $\sigma_s$ which denotes the fuzziness of the shell which also decreases very sharply at the VPTT following the expected ordering in $c$. Indeed, while at low temperatures $\sigma_s$ systematically decreases with increasing $c$, above the VPTT it converges to a common value of just a few nanometers for all microgels. This negligible interfacial width confirms a complete particle collapse and a loss of the distinct fuzzy-shell structure. The last reported parameter $\mu$, in panel (d), is related to the Flory solvency parameter $\nu$ through the relation $\mu=(1/\nu)-1$. From the value expected in good solvent at low $T$, $\mu\approx 0.66$, it increases with temperature, as expected, due to the worsening of the solvent condition. 

\begin{figure}[th!]
    \centering
    \includegraphics[width=1\linewidth]{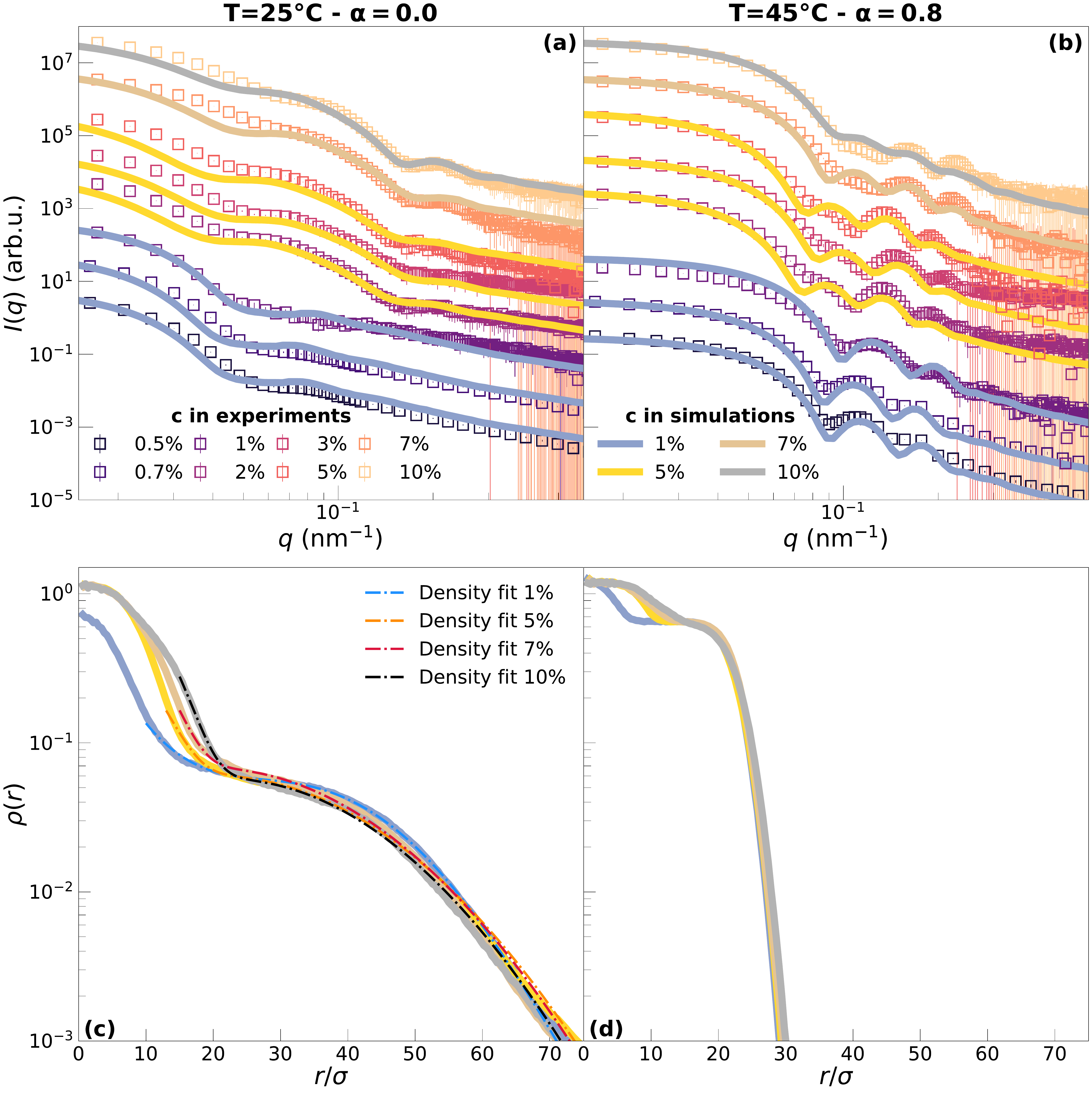}
    \caption{Experimental SAXS intensities (symbols) of EGDMA-crosslinked microgels synthesized with 1.6 mM SDS overlaid with the form factors calculated for \textit{in silico} microgel models (solid lines) for (a) $T= 25\celsius$ ($\alpha=0.0$) and (b) $45\celsius$ ($\alpha=0.8$).
    The numerical curves correspond to the best description of the experimental data where the $c$ content is used as adjustable parameter with respect to the nominal one;   corresponding radial density profiles, $\rho(r)$, of the \textit{in silico} microgels at the respective effective temperatures $\alpha=0.0$ (c) and $\alpha=0.8$ (d). The dashed lines in panel (c) represent fits of the profiles according to Eq.~\ref{eq:Rho_likos}.}
    \label{fig:EGDMA_T25-45_sim_SDS1.6mM}
\end{figure}

\noindent We finally compare SAXS data with numerical simulations  with $N=42000$ at varying crosslinker concentration, determining in this way the effective EGDMA content, which can differ from its nominal value, that best matches the experimental data.
In Figure~\ref{fig:EGDMA_T25-45_sim_SDS1.6mM} (a) and (b), we thus report the comparison between the calculated form factors for the \textit{in silico} microgels and the measured ones at $T=25\celsius$ and $T=45\celsius$, respectively. The mapping between the effective temperature $\alpha$ and real temperature $T$ 
is kept fixed, as described in Methods. Instead, the $q$-scale is adjusted for each simulated microgel by using a specific monomer size, which is reported in Table \ref{tab:sigma}.
In order to provide the best agreement, we allow $c$ to be a free parameter, in the sense that in the simulations, we choose the value of $c$ to ensure that the form factor of the corresponding microgel best reproduces the experimental one, irrespective of the experimental crosslinker density.
This is due to the fact that, as observed in the evolution of the form factor fit parameters reported in figure \ref{fig:star_fuzzysphere_parameters_SDS1.6mM}, results for $c\geq 2\%$ are all very similar to each other and do not follow a progressive evolution as it would be seen in the simulations. For completeness, the simulated form factor with varying nominal $c$ at low $T$ are reported in Fig.~\ref{fig:EGDMA_alpha00-08_sim_N42k}.
It is notable that the SAXS data for microgels with nominal $c\leq 1\%$ are all well described by the \textit{in silico} microgel with $c=1$\%. Instead, in order to adequately fit the experimental data for microgels with a nominal $c=2\%$, a substantially higher crosslinker content ($c=5$\%) is required in the simulations. This finding is consistent with the previous observation that the microgels separate into two distinct sets. The first set, with $c\leq 1$\%, exhibits star-like characteristics. The second set, with $c\geq 2$\%, shows a clear structural change, due to the larger crosslinker incorporation, which leads to a substantial increase of the microgel core size.

From a quantitative perspective, we find that a precise correspondence between the nominal crosslinker concentration used in the experimental synthesis and that used in the \textit{in silico} synthesis does not hold. Instead, the best match between the experimental and calculated form factors is achieved as follows: experimental data for $c$ of 0.5\%, 0.7\%, and 1\% correspond to a simulation with $c=1\%$; experimental data for $c=2\%$, 3\%, and 5\% correspond to $c=5\%$ in simulations;  finally experimental data for $c=7\%$ and 10\% can be identified with the nominal values 7\% and 10\% in simulations, respectively. These best fits are reported in Fig.~\ref{fig:EGDMA_T25-45_sim_SDS1.6mM}(a) and (b) for 25 and 45\celsius, respectively, denoting a rather good quantitative description of all experimental data with a single unifying numerical model.
Panels (c) and (d) of Figure~\ref{fig:EGDMA_T25-45_sim_SDS1.6mM} also show the corresponding radial density profiles, $\rho(r)$, directly calculated from simulations that best represent each experimental form factor. In particular, it is clear, as expected, that the core size increases with increasing $c$. Even at high $T$ a clear distinction between core and shell density remains. Interestingly, at low $T$, all the curves are compatible with the theoretical star behavior that can be expressed as proposed in Ref. ~\citenum{ruiz-francoMultiparticleCollisionDynamics2019} and detailed in the SI (eq.\ref{eq:Rho_likos}). Thus, a power-law decay is visible at intermediate distances followed by Gaussian decay at large ones, as shown by the fits also reported in Fig.~\ref{fig:EGDMA_T25-45_sim_SDS1.6mM}(c),  suggesting a star-like profile of the outer corona at all $c$, despite the core possesses a distinct, finite size. This confirms that the crosslinkers are mostly segregated inside the core and that the corona is predominantly made of free arms, independently on the different size and composition of the core when varying $c$.

\section{Conclusions}
In this work, we have investigated in detail the relationship between synthesis conditions and the resulting internal structure of thermoresponsive EGDMA-crosslinked microgels, with the aim of defining the optimal parameters for achieving a star-like architecture. First of all, we focused on the role played by the surfactant, commonly employed in the synthesis of microgels to target the desired size of the particles. While for BIS-crosslinked microgels, it is well-established that such a surfactant plays a minor role in the internal structure of the particles, the present
combined experimental and computational approach demonstrates that the presence of sodium dodecyl sulfate (SDS) is crucial for steering the particle morphology away from a conventional core-shell structure and toward the star-like regime.
Indeed, in the absence of SDS, even at a low nominal crosslinker concentration of 1\%, microgels develop a large, distinct core, as evidenced by the emergence of characteristic features in the SAXS form factor. This core is not solely composed of EGDMA but also incorporates a significant fraction of NIPAM monomers. This morphology was successfully captured by the fuzzycore-fuzzyshell model, which generalizes the previously established core-fuzzyshell model, used to fit microgels synthesized with surfactant, by allowing the core to have a fuzzy interface with the shell. Importantly, the same \textit{in silico} model previously developed to model EGDMA-crosslinked microgels synthesized with surfactant~\cite{ballinStarLikeThermoresponsiveMicrogels2025} can be directly applied to the microgels synthesized without SDS, by simply employing a sufficiently large number of monomers ($N=336000$). This pronounced size-dependence in both experiments and simulations  highlights the fundamental difference of the growth mechanism in the absence of surfactant with respect to BIS-crosslinked microgels. Furthermore, in the absence of SDS, microgels with $c > 2\%$ were not found to be stable after the synthesis, rather forming larger aggregates that had to be discarded from the present study.

\noindent Next, we verified that the addition of just a small amount of SDS (0.4 mM) in the synthesis is sufficient to restore star-like behavior for microgels with $c\leq 1\%$. This is evidenced by the form factor profile, again compatible with the core-fuzzyshell model, which is accompanied by a sharp volume phase transition and a high swelling ratio. These are the hallmarks of microgels with star-like architecture, making them promising for applications and for fundamental studies~\cite{papettiStarlikeMicrogelsVs2026}. 
These findings indicate that SDS is able to stabilize the growth of the core during the early stages of polymerization, preventing the increase of its size that would compromise the star-like structure. Moreover, the addition of 0.4 mM SDS enables the synthesis of stable microgels also at high EGDMA content, up to $c=10$\%. These particles have dimensions that are found to be comparable to those synthesized without surfactant (at low EGDMA content), yet possess structural features, including the form factor and swelling behavior, that closely resemble those of microgels prepared with a higher surfactant concentration (1.6 mM).

\noindent Finally, we investigated in detail the role played by crosslinker density at high SDS concentration, in order to establish the optimal regime for star-like behavior. We find that already for $c= 2\%$, the presence of a finite core becomes relevant, so that the internal structure is dominated by a dense, persistent core regardless of the amount of SDS used in the synthesis. In this regime, the particle architecture changes with increasing $c$, from a star-like to a core-shell morphology, both of which can be successfully described by the core-fuzzyshell star model.

\noindent In summary, the present study provides a comprehensive framework for the rational synthesis of EGDMA-crosslinked microgels. We have established that the star-like architecture, a desirable state for creating ultra-soft colloids, is achievable under two key conditions: a low molar fraction of EGDMA ($c\leq 1\%$) and the presence of even a small amount of surfactant to stabilize core growth. These findings not only bridge the gap between the fields of star polymers and microgels but also provide a practical synthesis guide for creating model soft particles with tunable interactions for fundamental studies and applications. We are confident that our results will stimulate scientific activity for utilizing EGDMA-crosslinked microgels in a wide range of studies, from the fundamental exploration of the physics of crowded environments and rheological properties of soft colloids, to the application as high-performance components in smart delivery systems and responsive materials.

\begin{acknowledgement}
We thank Marco Laurati, Jacopo Vialetto, Tommaso Papetti, Thomas Hellweg and Jan Skov Pedersen for stimulating discussions. 
We acknowledge the European Synchrotron Radiation Facility (ESRF) for provision of synchrotron radiation facilities  under proposal number SC-5593 \cite{DatiESRF}. EB and EZ acknowledge financial support by Progetto Co-MGELS funded by the European Union - NextGeneration EU under the National Recovery and Resilience Plan (PNRR) Mission 4 “Istruzione e Ricerca” - Component C2 - Investment 1.1 - "Fondo PRIN", Project code PRIN2022PAYLXW Sector PE11, CUP B53D23008890006. FB, SS and EZ also acknowledge financial support from INAIL, project MicroPad (BRiC 2025, ID 66) and from ERC POC project MICROSENS (grant agreement no.101157420).  We gratefully acknowledge the CINECA award under the ISCRA initiative, for the availability of high-performance computing resources and support.

\end{acknowledgement}

\section*{Supporting Information}

We separately provide a Supporting Information file containing additional data on the microgel characterization .
\bibliography{biblio}

\newpage
\begin{suppinfo}
\setcounter{figure}{0}
\setcounter{table}{0}
\setcounter{page}{1}
\renewcommand{\theequation}{S\arabic{equation}}
\renewcommand{\thefigure}{S\arabic{figure}}
\renewcommand{\thetable}{S\arabic{table}}

Figure~\ref{fig:DatiSAXS_SDS0.0mM} reports the measured intensity curves for the two microgels synthesized without added surfactant and with $c=1\%$ (a) and 2\% (b) in a temperature range between 25 and 45 \celsius.

\begin{figure}[H]
    \centering
    \includegraphics[width=1\linewidth] {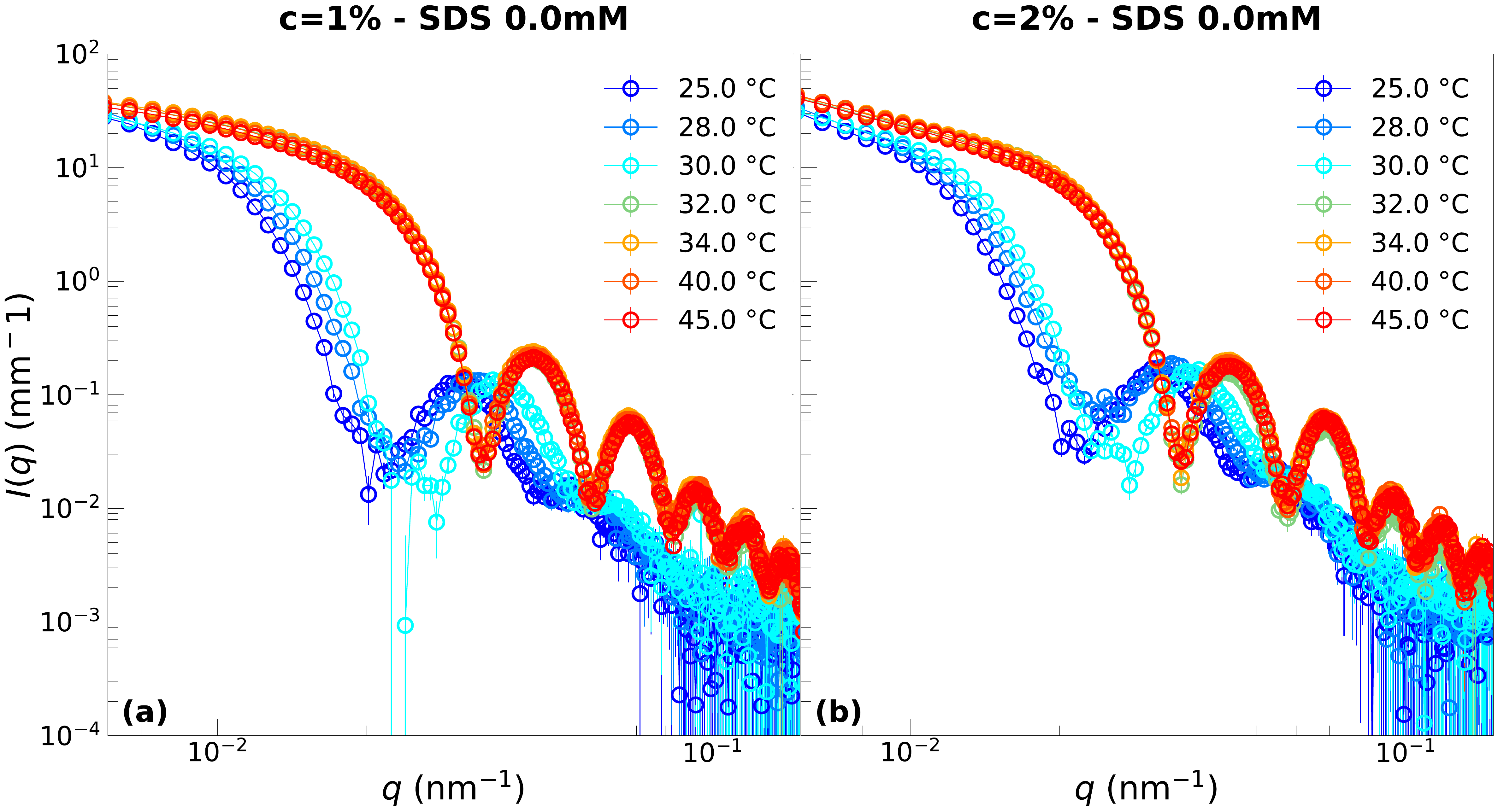}
    \caption{Intensity curves $I(q)$ measured via USAXS in a temperature range between $25\celsius$ and $45\celsius$ for samples synthesized without added surfactant and with $c=1\%$ (a) and $c=2\%$ (b).}
    \label{fig:DatiSAXS_SDS0.0mM}
\end{figure}

\noindent Figure~\ref{fig:EGDMA2_fit_sim_rho_SDS0.0mM} displays the USAXS characterization for the microgel synthesized with $c=2\%$ without added surfactant. In particular, Fig.~\ref{fig:EGDMA2_fit_sim_rho_SDS0.0mM}(a) shows the form factors measured at temperatures ranging from $25 \celsius$ to $45 \celsius$ together with fits through Eq.~\ref{eq:model_fc-fs-lorentz}, while  Fig.~\ref{fig:EGDMA2_fit_sim_rho_SDS0.0mM}(b) reports the density profiles calculated according to Eq.~\ref{eq:rho} using the best fit parameters of curves reported in panel (a).
Next, Figs.~\ref{fig:EGDMA2_fit_sim_rho_SDS0.0mM}(c) and (d) compare the experimental form factors and radial density profiles, respectively, with those calculated from simulations of an \textit{in silico} microgel with the same nominal cross-linker concentration ($c=2\%$) and $N=336000$. In order to rescale simulations on experimental curves we shifted the calculated form factors using a monomer size $\sigma=2.90$ nm. 
Also in this case, as for $c=1\%$ reported in the main text, we find excellent agreement between the experiment and simulations.
\begin{figure}[H]
    \centering
    \includegraphics[width=1\linewidth] {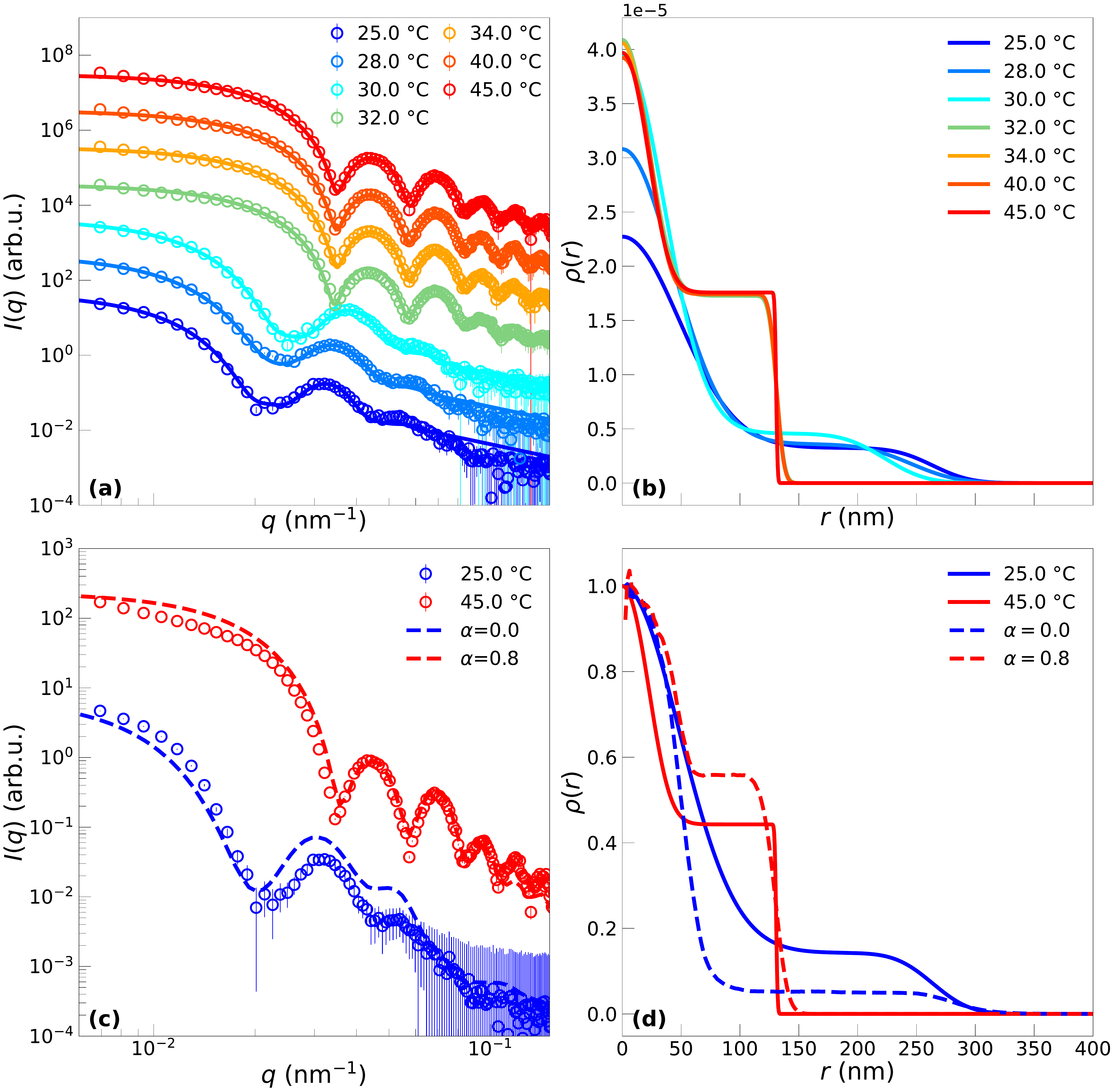}
    \caption{(a) Fits of the USAXS data for the microgel with $c=2\%$ synthesized without SDS using the fuzzycore-fuzzyshell model reported in equation \ref{eq:model_fc-fs-lorentz} across different temperatures. To improve visualization, the curves were arbitrarily offset in the \textit{y}-direction. (b) Radial density profiles, $\rho(r)$, obtained via Fourier transform of the square root of equation \ref{eq:fuzzycore-fuzzyshell_model} with the best-fit parameters of the curves shown in panel (a) as described in equation \ref{eq:rho}.  Curves are normalized such that the spherical integral of the $\rho(r)$ equals one. (c) Comparison of the experimental form factors with those calculated on \textit{in silico} microgels with the same nominal crosslinker concentration.  (d) Direct comparison between the experimentally derived $\rho(r)$ profiles and those from the simulations. Curves are normalized such that $\rho(0)=1$.
    }
    \label{fig:EGDMA2_fit_sim_rho_SDS0.0mM}
\end{figure}
\noindent We find that the microgels synthesized with $c=2\%$ are very similar to those synthesized with $c=1\%$, as confirmed also by the values of the fit parameters of the form factor curves reported in figure \ref{fig:fuzzycore_fuzzyshell_parameters_SDS0.0mM}.
For both samples, the core radius ($r_c$, panel (a)) remains constant across the temperature range, while the core fuzziness ($\sigma_c$, panel (b)), shell thickness ($t=r_s-r_c-2\sigma_c$, panel (c)), and shell fuzziness ($\sigma_s$, panel (d)) all decrease at the VPTT, with the fuzziness parameters showing similar trends for both compositions.

\begin{figure}[H]
    \centering
    \includegraphics[width=1\linewidth] {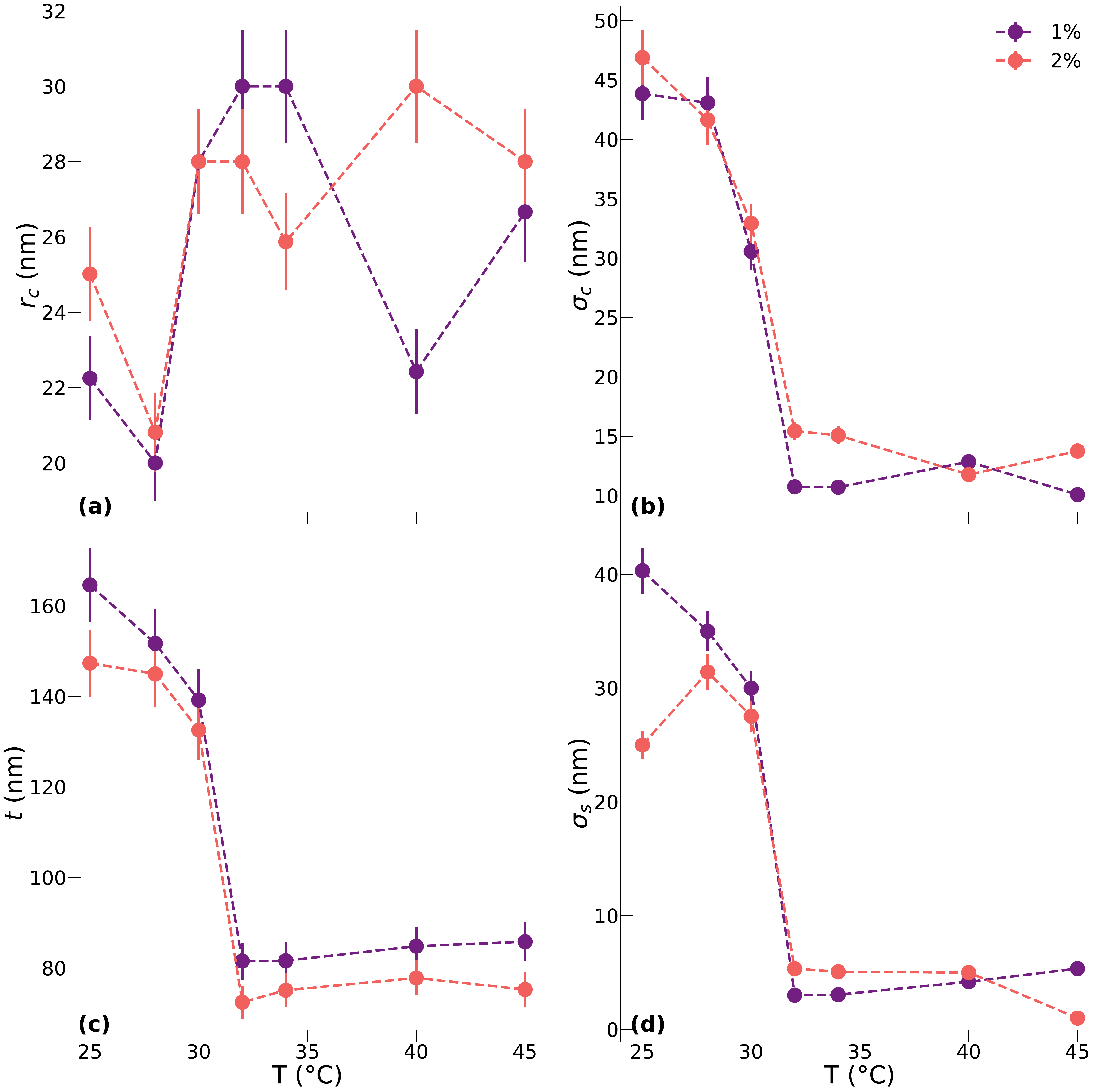}
    \caption{Fit parameters of the scattering intensity curves presented in panel (a) of figures \ref{fig:EGDMA1_fit_sim_rho_SDS0.0mM} and \ref{fig:EGDMA2_fit_sim_rho_SDS0.0mM} for samples synthesized without SDS with $c = 1\%$ and 2\% as a function of temperature: (a) radius of the core $r_c$; (b) fuzziness of the core $\sigma_c$; (c) thickness of the shell $t=r_s-r_c-2\sigma_c$; (d) fuzziness of the shell $\sigma_s$.}
\label{fig:fuzzycore_fuzzyshell_parameters_SDS0.0mM}
\end{figure}
\noindent Figure~\ref{fig:DatiSAXS_SDS0.4mM} reports the scattering curves for samples synthesized with 0.4 mM SDS at $c= 1\%$ (a) and $10\%$ (b), measured across a temperature range from $25\celsius$ to $45\celsius$.

\begin{figure}[H]
    \centering
    \includegraphics[width=1\linewidth] {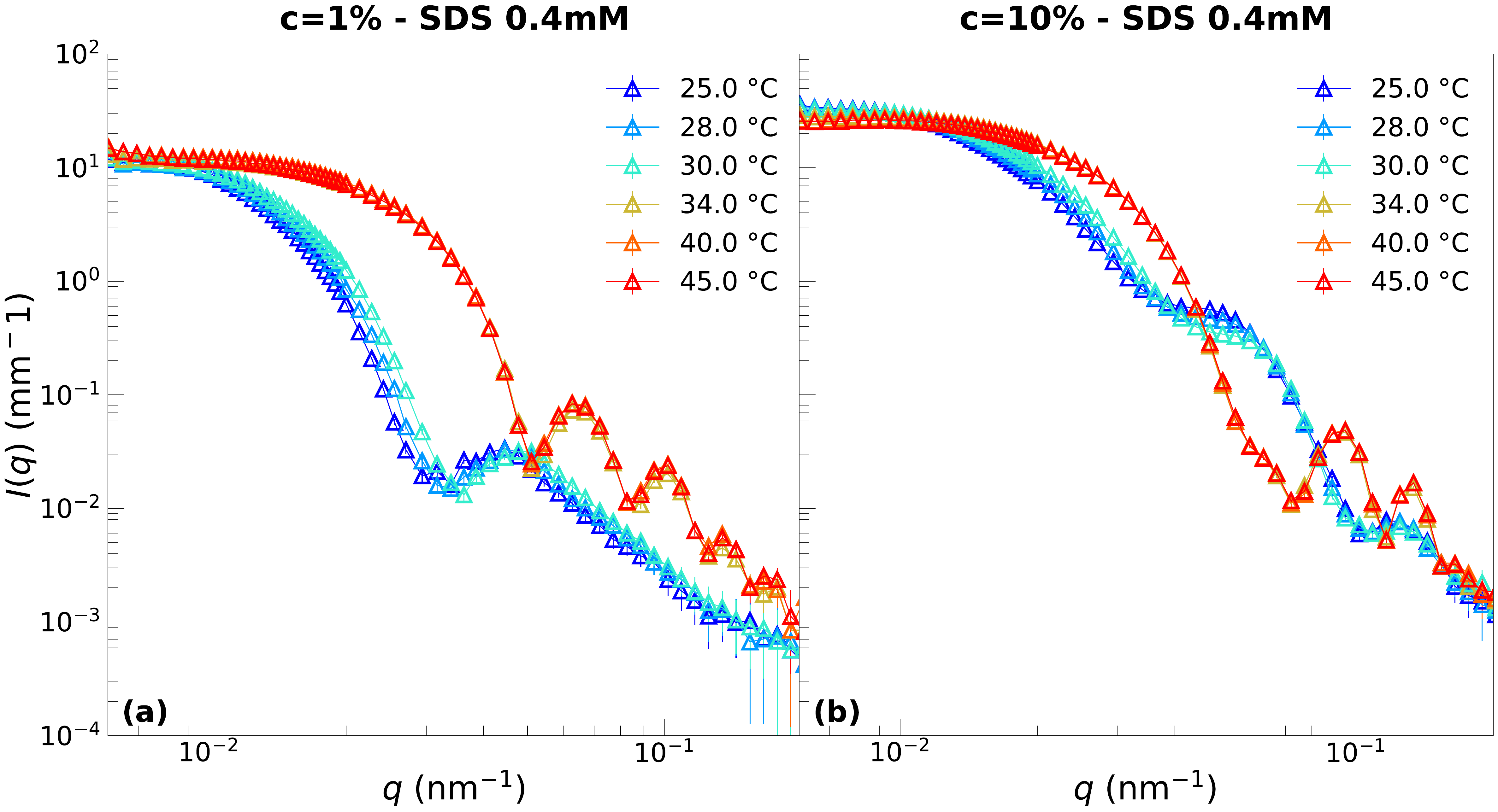}
    \caption{Intensity curves $I(q)$ measured via USAXS in a temperature range between $25\celsius$ and $45\celsius$ for samples synthesized with 0.4 mM of SDS and with $c=1\%$ (a) and $c=10\%$ (b).}
    \label{fig:DatiSAXS_SDS0.4mM}
\end{figure}
\noindent Figure~\ref{fig:star_fuzzysphere_parameters_SDS0.4mM} displays the trend of the best-fit parameters as a function of temperature obtained from applying the core-fuzzy shell star model (Eq.~\ref{eq:model_c-fs-star}) to the samples synthesized with 0.4 mM SDS. The corresponding fitted curves are shown in Figure~\ref{fig:EGDMA1-10_fit_SDS0.4mM}.

\begin{figure}[H]
    \centering
    \includegraphics[width=1\linewidth] {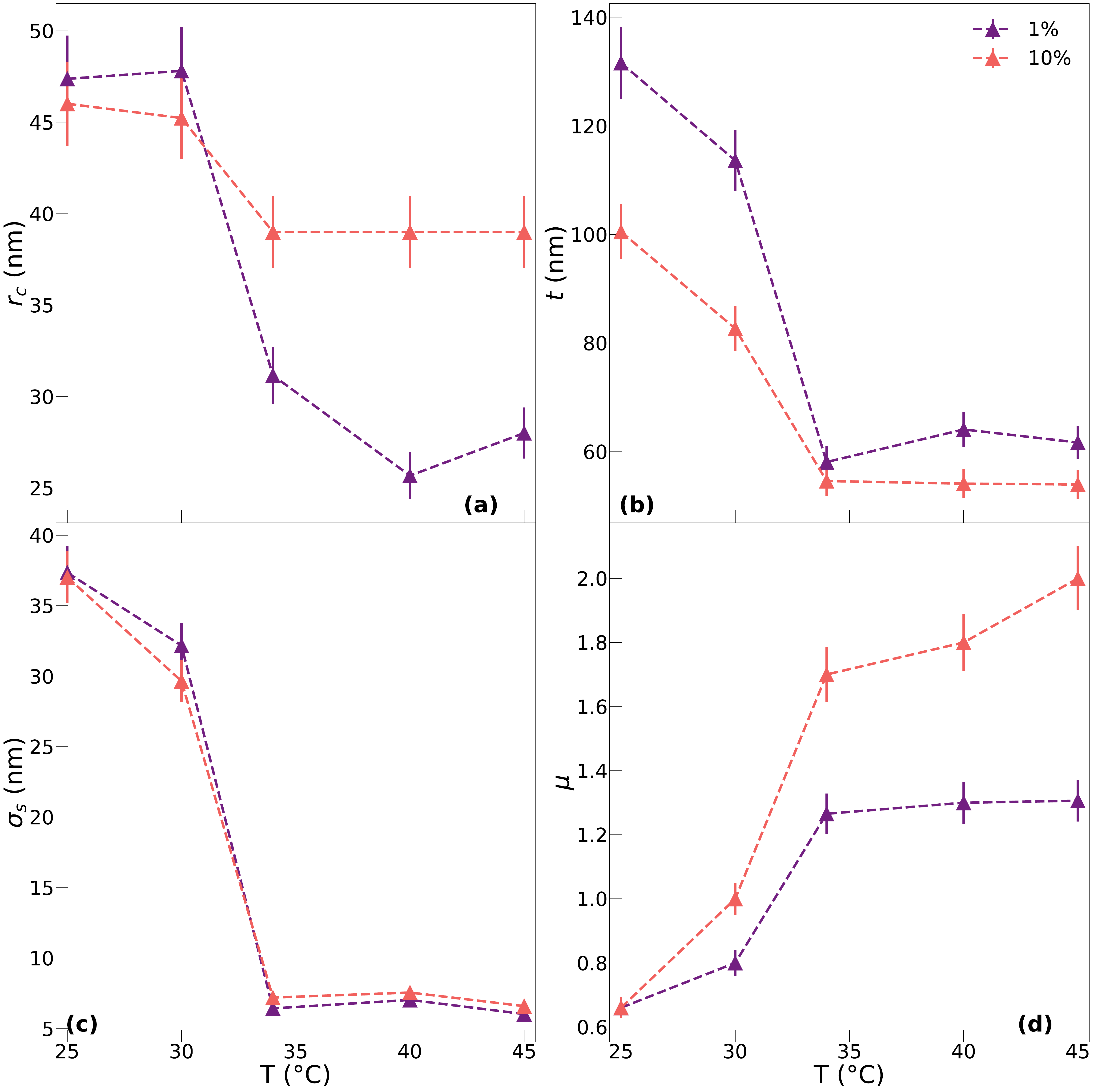}
    \caption{Fit parameters of the scattering intensity curves presented in panels (a) and (b) of Fig.~\ref{fig:EGDMA1-10_fit_SDS0.4mM} for samples synthesized with 0.4 mM of SDS with $c= 1\%$ and 10\% as a function of temperature: (a) radius of the core $r_c$; (b) thickness of the shell $t$; (c) fuzziness of the shell $\sigma_s$; (d) $\mu = (1/\nu)-1$, with $\nu$ the Flory solvency parameter.}
    \label{fig:star_fuzzysphere_parameters_SDS0.4mM}
\end{figure}

\noindent Figure~\ref{fig:all_fit_SDS1.6mM} reports the form factors in a temperature range between $25\celsius$ and $45\celsius$ fitted through the core - fuzzy shell star model (equation \ref{eq:model_c-fs-star}) for microgels synthesized with different content of EGDMA and 1.6 mM of SDS.
\begin{figure}[H]
    \centering
    \includegraphics[width=0.9\linewidth]{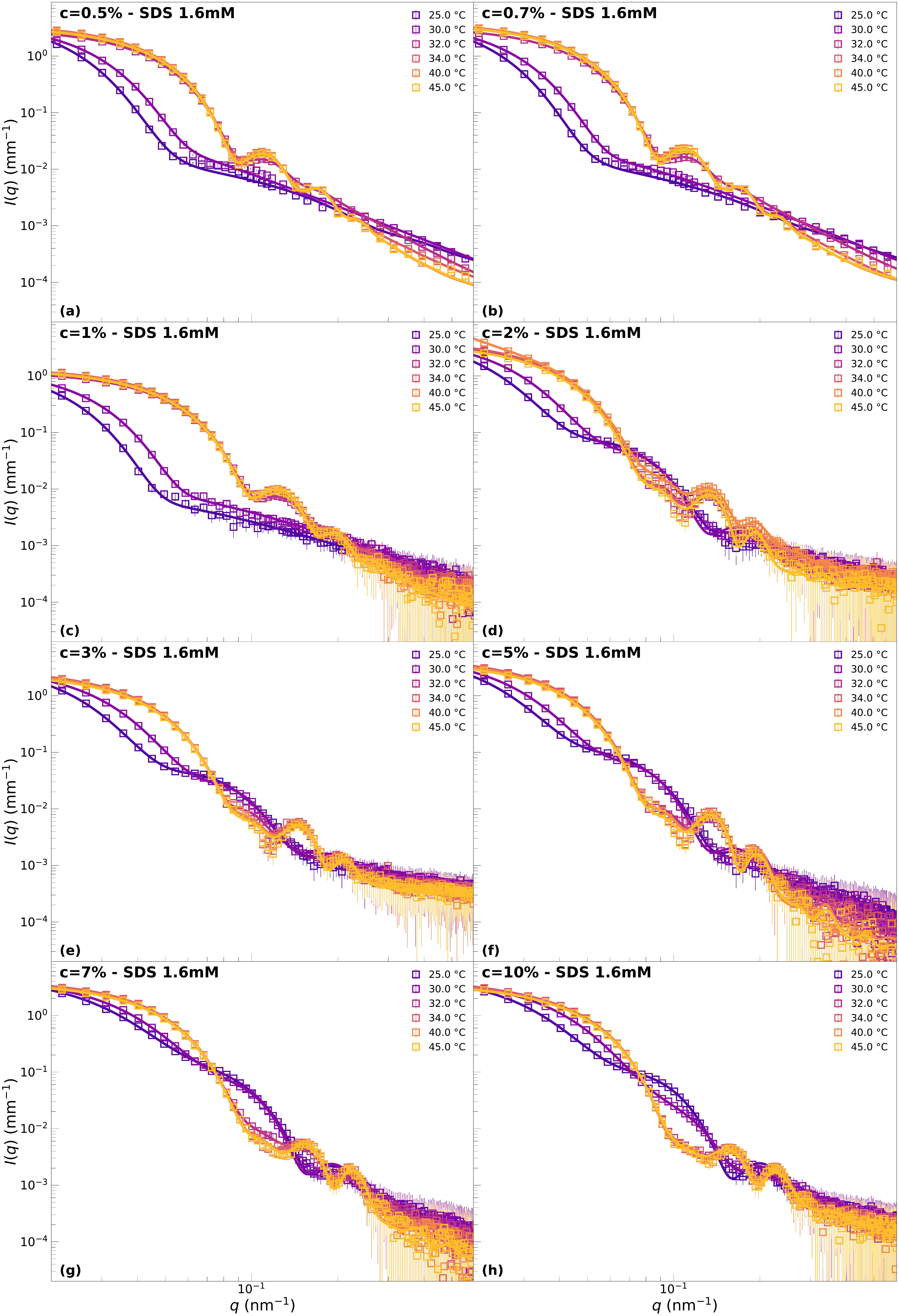}
    \caption{Form factors fitted through the core-fuzzyshell star model (Eq.~\ref{eq:model_c-fs-star}) at different temperatures in the range between $25\celsius$ and $45\celsius$ for all the samples synthesized with 1.6 mM of SDS.}
    \label{fig:all_fit_SDS1.6mM}
\end{figure}

\noindent Table~\ref{tab:sigma} reports the monomer sizes used to rescale simulations with $N=42000$ on the experimental form factors of microgels synthesized with 1.6 mM of SDS at different $c$ displayed in Fig. \ref{fig:EGDMA_T25-45_sim_SDS1.6mM} (a) and (b).

\begin{table}[]
    \centering
    \begin{tabular}{|cc|}
    \hline
        $c$ (\%) &  $\sigma$ (nm)\\ \hline
        0.5 & 2.27 \\
        0.7 &2.38 \\
        1 & 2.08\\
        2 & 2.94\\
        3 & 2.63\\
        5 & 2.86\\
        7 & 2.38\\
        10 &2.33 \\ \hline
        
    \end{tabular}
    \caption{Monomer sizes used to rescale simulations with $N=42000$ on the experimental form factors of microgels synthesized with 1.6 mM of SDS.}
    \label{tab:sigma}
\end{table}

\noindent Figure~\ref{fig:EGDMA_alpha00-08_sim_N42k} reports the calculated form factors for \textit{in silico} microgels with N=42000 at all EGDMA concentrations used experimentally for the low-temperature state ($\alpha = 0.0$).
\begin{figure}[H]
    \centering
    \includegraphics[width=0.6\linewidth] {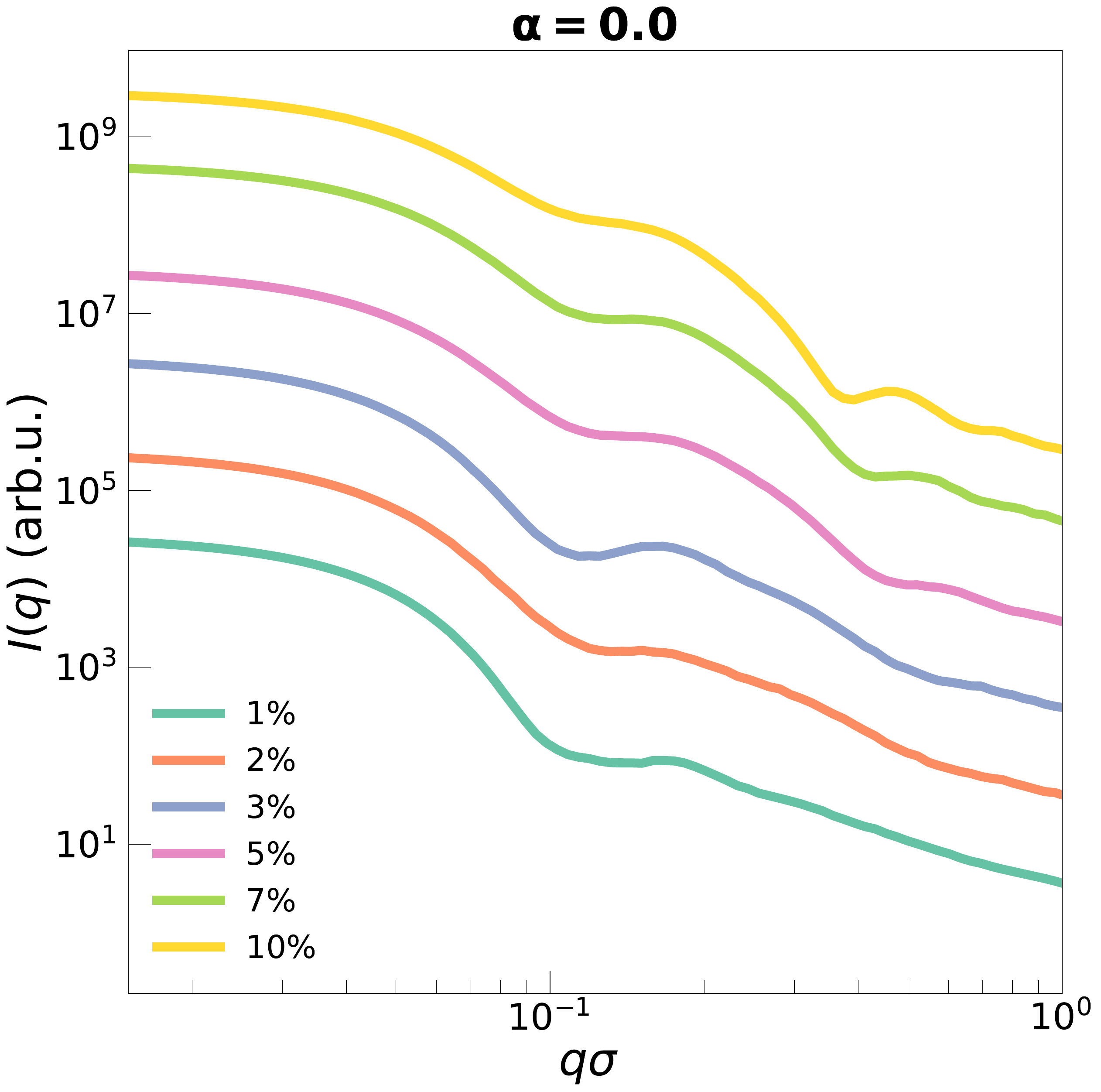}
    \caption{Calculated form factors of  \textit{in silico} microgels with different EGDMA content in the low temperature state ($\alpha=0.0$) and N=42000.}
    \label{fig:EGDMA_alpha00-08_sim_N42k}
\end{figure}

\noindent Table~\ref{tab:parameters_rho} reports the best fit parameters of the density profiles displayed in Figure \ref{fig:EGDMA_T25-45_sim_SDS1.6mM}(c) according to the predicted star behavior~\cite{ruiz-francoMultiparticleCollisionDynamics2019}, which reads as:
\begin{equation}
\label{eq:Rho_likos}
\rho^{star}\left(r\right)\sim A_{1}\,r^{-4/3}\,f_{b}\left(r,r_{b}\right)+ A_{2}\left[1-f_{b}\left(r,r_{b}\right)\right]\exp\left[-\left(\frac{r-r_{1}}{r_{2}} \right)^{2}\right],
\end{equation}
with $f_{b}\left(r,r_{b}\right)= \exp\left[-\left(\frac{r}{r_{b}}\right)^{4}\right]$ a bridge function between the two regimes.

\begin{table}[H]
    \centering
    \begin{tabular}{|cccccc|}
    \hline
    $c$ (\%)& $A_1$ & $r_b$ $[\sigma]$& $A_2$& $r_1$ $[\sigma]$& $r_2$ $[\sigma]$\\ \hline
    1&     3.45& 19.55&0.06& 23.15 &25.31\\
    5&6.85& 16.01& 0.06 &20.52 &26.29\\
    7&11.11 &15.59 &0.067& 19.73 &25.94\\
    10& 19.41 &16.17 &0.057& 21.99 &24.70\\ \hline
    \end{tabular}
    \caption{Best fit parameters according to Eq.~\ref{eq:Rho_likos} of the calculated density profiles on \textit{in silico} microgels with $N=42000$ and different crosslinker contents. The lengths are given in units of the  monomer size in simulations $\sigma$.}
    \label{tab:parameters_rho}
\end{table}

\end{suppinfo}

\end{document}